\documentclass[fleqn,usenatbib]{mnras}

\usepackage{newtxtext,newtxmath}

\usepackage[T1]{fontenc}
\usepackage[percent]{overpic}
\usepackage{times}
\usepackage{graphicx}
\usepackage{amsmath}

\usepackage{mathabx}
\usepackage{subfigure}
\usepackage{natbib}
\usepackage{txfonts}
\usepackage{xcolor}
\usepackage[utf8]{inputenc}
\usepackage[overload]{empheq}
\usepackage{verbatim}
\usepackage{CJKutf8}
\usepackage{CJK}

\DeclareRobustCommand{\VAN}[3]{#2}
\let\VANthebibliography\thebibliography
\def\thebibliography{\DeclareRobustCommand{\VAN}[3]{##3}\VANthebibliography}

\usepackage{graphicx}	
\usepackage{amsmath}	

\title[Photometric Re-calibration of VPHAS+ $u$-band Data]{Photometric Re-calibration of VPHAS+ $u$-band Photometry with the Stellar Colour Regression Method and Gaia DR3}

\author[B.-Q. Chen et al.]{
Bing-Qiu Chen,$^{1}$\thanks{E-mail: bchen@ynu.edu.cn.}
Hai-Bo Yuan$^{2,3}$\thanks{E-mail: yuanhb@bnu.edu.cn.}
and Bo-Wen Huang$^{2,3}$
\\
$^{1}$South-Western Institute for Astronomy Research, Yunnan University, Kunming, 650500, P.\,R.\,China\\
$^{2}$Institute for Frontiers in Astronomy and Astrophysics, Beijing Normal University, Beijing 102206, P.\,R.\,China\\
$^{3}$Department of Astronomy, Beijing Normal University, Beijing, 100875, P.\,R.\,China
}

\date{Accepted XXX. Received YYY; in original form ZZZ}

\pubyear{2024}

\begin{document}
\label{firstpage}
\pagerange{\pageref{firstpage}--\pageref{lastpage}}
\maketitle

\begin{abstract}
The $u$ band magnitude is vital for determining stellar parameters and investigating specific astronomical objects.  However, flux calibration in the $u$ band for stars in the Galactic disk presents significant challenges. In this study, we introduce a comprehensive re-calibration of $u$-band photometric magnitudes of the VPHAS+ Data Release 4 (DR4), employing the Stellar Colour Regression (SCR) technique. By leveraging the expansive set of XP spectra and $G_{\rm BP}$ photometry from Gaia Data Release 3 (DR3), as well as the individual stellar extinction values provided by the literature, we have obtained precise model magnitudes of nearly 3 million stars. Our analysis identifies systematic magnitude offsets that exhibit a standard deviation of 0.063\,mag across different observational visits, 0.022\,mag between various CCDs, and 0.009\,mag within pixel bins. We have implemented precise corrections for these observational visits, CCD chips, and pixel bins-dependent magnitude offsets. These corrections have led to a reduction in the standard deviation between the observed magnitudes and the model magnitudes from 0.088\,mag to 0.065\,mag, ensuring that the calibrated magnitudes are independent of stellar magnitude, colour, and extinction. The enhanced precision of these magnitudes substantially improves the quality of astrophysical research and offers substantial potential for furthering our understanding of stellar astrophysics.
\end{abstract}




\section{Introduction}

In recent years, numerous wide-field photometric surveys have been conducted \citep{York2000,DES2005,Skrutskie2006,Wright2010,Chambers2016,Gaia2016,Zou2017,Onken2019,Ivezic2019}, producing revolutionary impacts on the advancement of astronomy. Uniform and high-precision flux calibration, which ensures a high degree of consistency in the flux measurements between targets that are widely separated in the sky and guarantees that these measurements do not vary with changes in the telescope's working environment, detector position, or time of observation, is one of the key factors determining the success of wide-field surveys \citep{Kent2009,Huang2022}. The precision of flux calibration sets the lower limit on the error in stellar brightness measurements, which is significant for the determination of stellar photometric parameters, the selection of special astronomical objects, and the study of the structure of the Milky Way. Therefore, for the successful execution of wide-field photometric surveys, the undertaking of high-precision flux calibration is of utmost importance.

In traditional astronomical observations, absolute flux calibration is commonly performed using standard stars that are known to have a good internal consistency of brightness \citep{Landolt1992,Yan2000,Zhou2001}. However, due to a shortage of standard stars, the Forbes effect, the colour-dependent atmospheric extinction effect, and flat-field precision, the classical standard star method has difficulties to meet the milli-magnitude level calibration precision required by contemporary wide-field photometric surveys \citep{Huang2022}. In recent years, astronomers have developed a series of new flux calibration methods for wide-field photometric surveys. These methods typically focus only on the relative flux calibration process, with the absolute zero point still being provided by observing standard stars or calibrated to specific standard photometric systems. These include the Uber-calibration method \citep{Padmanabhan2008}, the Hyper-calibration method \citep{Finkbeiner2016}, the Forward Global Calibration Method \citep{Burke2018}, and the XP Synthetic Photometry method \citep{Xiao2023,Xiao2023b}; as well as the Stellar Locus Regression method \citep{High2009}, Stellar Locus method \citep{Sanjuan2019}, and Stellar Colour Regression (SCR) method \citep{Yuan2015,HuangYuan2022}, among others.

Through the application of these methods, we have now achieved typical flux calibration precision of a few milli-magnitude in ground-based optical bands \citep{Huang2022}. However, the calibration precision in the near-ultraviolet $u$ band, especially for the $u$ band photometry in the Galactic disk, is still significantly larger than that in the optical bands. This is due to factors such as more significant atmospheric extinction effects in the $u$ band, more sensitive to stellar atmospheric parameters, lower detector response, fainter stellar magnitudes, insufficient numbers of standard stars, and more serious interstellar extinction effects in the Galactic disk. Nevertheless, the $u$ band is a very important part of stellar parameter determination and the identification of special astronomical objects. Previous work has attempted to accurately calibrate the $u$ band photometry for stars at high Galactic latitudes using stellar atmospheric parameters. \citet{Yuan2015} and \citet{HuangYuan2022} have refined the $u$ band photometric calibration for the SDSS Stripe 82, achieving a precision of approximately 5\,mmag in $(u-g)$ colour and $u$ magnitude, respectively. In a similar vein, \citet{Huang2021} managed to re-calibrate the $u$ band data for the SkyMapper Southern Survey at high Galactic latitudes with a precision of up to 1\,per\,cent. Furthermore, the $u$ band photometry for the J-PLUS and S-PLUS surveys have been re-calibrated by \citet{Xiao2023J} and \citet{Xiao2023S}, respectively, attaining calibration precision around of 5\,mmag.

The VST Photometric H$\alpha$ Survey (VPHAS+; \citealt{Drew2014}) uses VST/OmegaCAM to survey the Southern Galactic Plane and Bulge in $u, g, r, i$, H$\alpha$ down to magnitudes greater than 20. The $u$ band photometric data from VPHAS+ are essential for astronomical research, as they allow for the precise determination of stellar parameters near the Galactic centre and the study of the metallicity of the inner Galactic disk and the bulge. They are also crucial for the selection of OB stars in the inner Galactic disk. \citet{Chen2019a}, using the VPHAS+ DR2 $u$ band data, identified nearly 27,000 candidates for O-type and early B-type stars, but due to the poor photometric precision, most of these sources were false positives. After validation with Gaia Data Release (DR2) data \citep{GaiaDR2}, of the nearly 27,000 candidates, only 6,858 were confirmed as candidates for O-type and early B-type stars.

The VPHAS+ observations concluded in 2018, and VPHAS+ Data Release (DR4) released catalogues of objects based on all observational data, including aperture magnitudes in the $u$ band for stars. The main goal of this paper is to assess the $u$ band photometric precision of stars in this VPHAS+ final data release and attempt to further calibrate the photometry. The structure of this paper is as follows: in Section 2, we introduce the data used; in Section 3, we present the methods employed. Our results are showcased and discussed in Section 4, and concluded in Section 5.

\begin{figure*}
\centering
	\includegraphics[height=0.78\columnwidth]{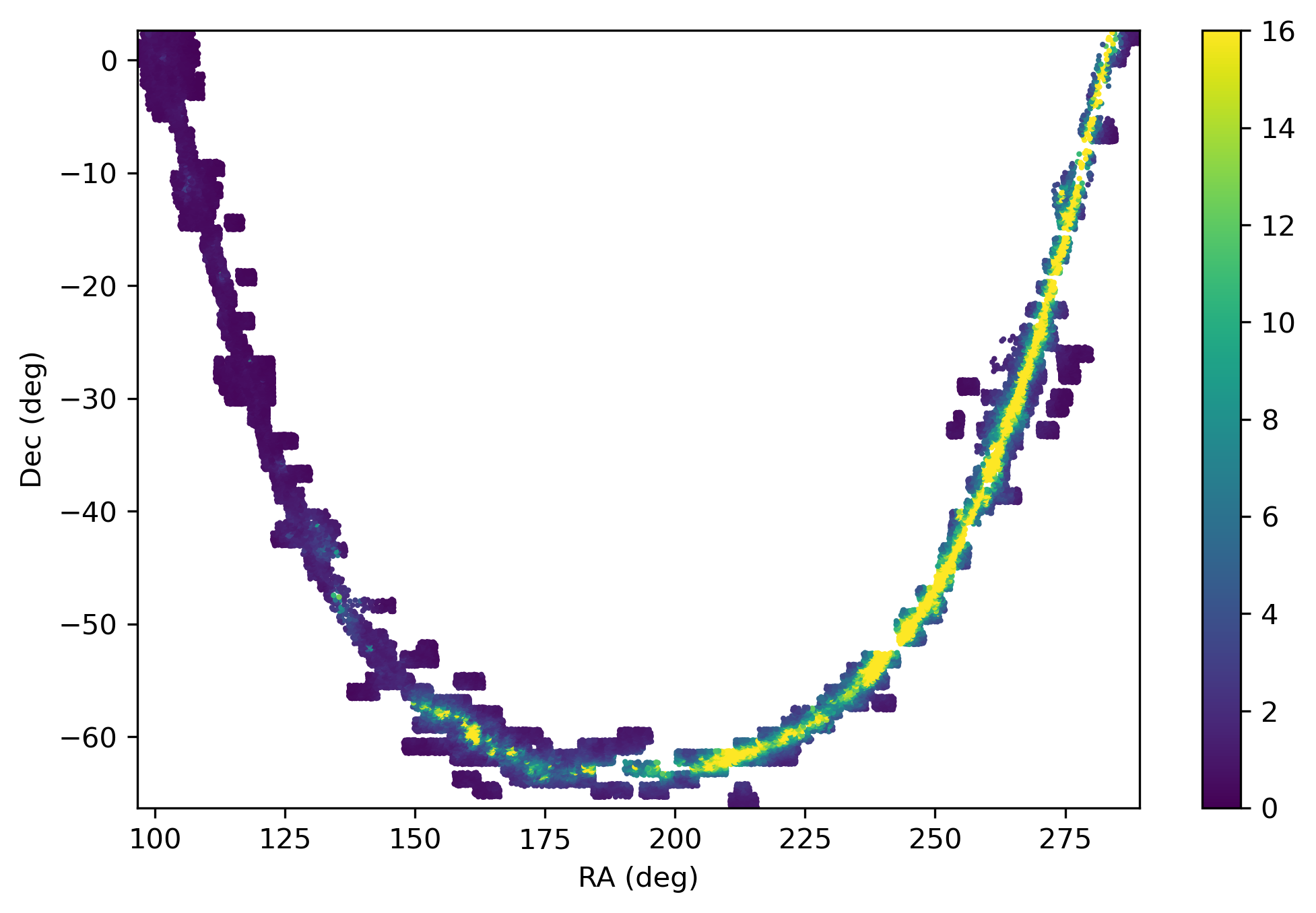}
 	\includegraphics[height=0.78\columnwidth]{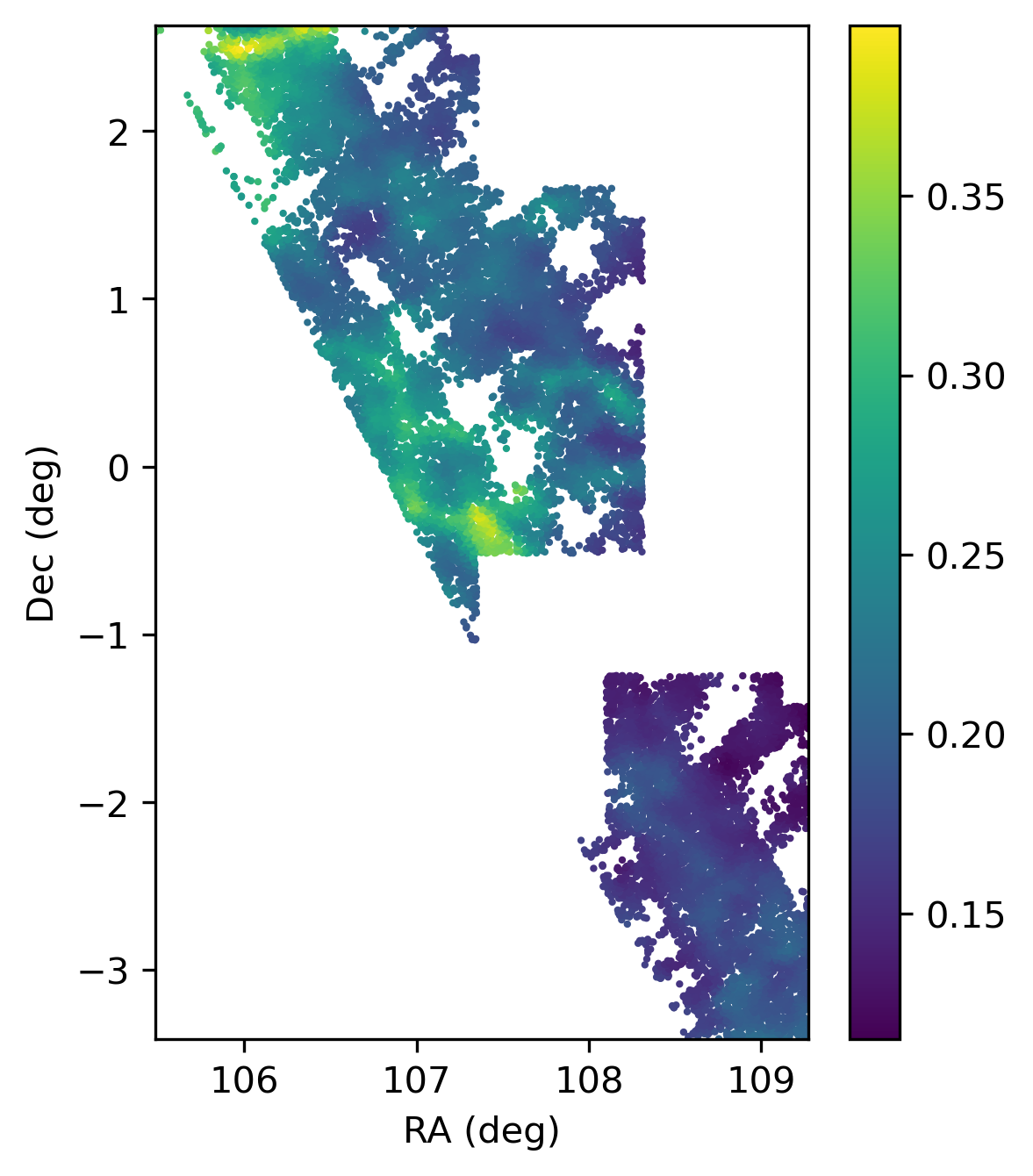}
    \caption{Spatial distribution of the Calibration Sample (left) and the Control Sample (right), respectively. In the Calibration Sample, only every one in 100 stars is shown for clarity. The colour scales represent the total LOS extinction derived from \citet{Planck2014} in the respective regions.}
    \label{spadist}
\end{figure*}

\section{Data}

\subsection{VPHAS+}
In this work, we present the utilization of the VPHAS+ DR4. This release encompasses observations conducted from 2015 to 2018. The data set includes reduced images and unstacked single-band source lists. The reduced images retain the original format of 32-CCD OmegaCam pawprints, covering a one-square-degree tile, and are not co-added. The single-band catalogues are derived from these images by the Cambridge Astronomical Survey Unit (CASU) pipeline.

For this study, we predominantly concentrate on $u$-band photometry, archiving 1,544 catalogues including $u$-band magnitudes. The number of objects detected in each catalogue varies significantly, with counts ranging from approximately ten thousand to hundreds of thousands, depending on the observed sky regions. Astrometric calibration was conducted with reference to the Two Micron All Sky Survey (2MASS) catalogue \citep{Skrutskie2006}, while photometric zeropoint calibration was performed nightly using standard stars. The astrometric precision within the data processing pipeline has a typical mean RMS error of 70 -- 80\,mas with respect to 2MASS. The source fluxes obtained were subjected to an illumination correction by cross-referencing with the AAVSO Photometric All-Sky Survey (APASS; \citealt{Henden2009}) stellar photometry, which, in turn, further refined the photometric zero-points. Reddening corrections were not applied to the data. The photometric calibration for the $u$ band is, however, relatively uncertain. The Vega $u$ band zeropoints are typically overestimated by about 0.3\,mag, and the AB $u$ band zeropoints are extrapolated from longer wavelength bands, which can be unreliable under various atmospheric conditions\footnote{See the VPHAS+ DR4 release description via \url{https://www.vphasplus.org/vphasplus-dr4-release-description.pdf}}. 

\subsection{Gaia}

For further refinement of the VPHAS+ DR4 $u$ band photometric calibration, we incorporate data from Gaia Data Release 3 (DR3; \citealt{GaiaDR3}). We utilize the high-precision $G_{\rm BP}$ band photometric data and the low-resolution (BP/RP) XP spectra from Gaia DR3. The Gaia $G_{\rm BP}$ band spans wavelengths from 330 to 680\,nm, providing photometry for approximately 1.5 billion stars. The calibration errors for the $G_{\rm BP}$ band are generally less than 1\,mmag, as per the overall trend, except for very blue and bright sources \citep{Riello2021,Yang2021}. We have corrected the Gaia $G_{\rm BP}$ magnitudes following the methodology proposed by \citet{Yang2021}.

Additionally, Gaia DR3 offers XP spectra for about 220 million stars \citep{Carrasco2021,GaiaDR3}. Rather than traditional pixelized wavelength-space measurements, the Gaia DR3 XP spectra are provided as sets of coefficients corresponding to 55 orthonormal Hermite functions for each of BP and RP spectra, resulting in a 110-dimensional coefficient vector for each stellar spectrum. For the purposes of our analysis, we require the spectral data in wavelength space to effectively exclude the influence of interstellar dust extinction. We have therefore transformed the Gaia DR3 XP spectra coefficients into wavelength space, utilizing the corrected spectra provided by \citet{Huang2024}. These spectra have been resampled to a uniform wavelength grid spanning 392 to 992\,nm, with a resolution of 10\,nm, aligning with the methodology of \citet{Zhang2023}. We note that the $u$ band encompasses 320 -- 390\,nm, suggesting the utility of XP spectrum data below 390\,nm. However, our decision to limit the spectrum range is driven by two primary considerations. First, the Gaia XP spectra display considerable noise near the boundaries of the BP and RP bands. The spectral data below 390\,nm are characterized by elevated uncertainty levels, which could adversely affect the precision of our recalibration efforts. Second, we have incorporated the extinction values and extinction law from \citet{Zhang2023}, which are specifically defined for the wavelength range of 392 to 992\,nm. This alignment enables direct comparison with their established extinction curve, which we utilize to ascertain the intrinsic stellar spectra. Despite omitting the spectra at lower wavelengths, the data from 392 - 992\,nm are sufficient to accurately determine the intrinsic colours of the stars under study.

Given that the VPHAS+ survey areas are situated at low Galactic latitudes with high extinction values, the commonly used two-dimensional extinction maps \citep{Planck2014} tend to overestimate stellar extinction values in these regions. To address this, we adopt the stellar extinction values $E0$ derived by \citet{Zhang2023}. They developed a forward model to estimate stellar atmospheric parameters (e.g., $T_{\rm eff}$, $\log g$, and $[\text{Fe}/\text{H}]$), distances, and extinctions for 220 million stars with XP spectra from Gaia DR3. They also provided an estimate of the optical extinction curve at the resolution of the XP spectra. Utilizing the extinction values and curve from \citet{Zhang2023}, we have obtained the intrinsic XP spectra for the individual stars.

\subsection{Sample selection}

We cross-match the VPHAS+ DR4 $u$-band photometric catalogue with the Gaia DR3 within a matching radius of 1 arcsecond. To define our `Calibration Sample' for flux calibration, we apply the following selection criteria:
\begin{enumerate}
  \item The source type from VPHAS+ is classified as a star with \texttt{MERGEDCLASS = -1};
  \item The default VPHAS+ $u$-band aperture-corrected magnitude is between 12 and 21.5\,mag, and the magnitude errors are smaller than 0.08\,mag;
  \item The Gaia \texttt{phot\_bp\_rp\_excess\_factor} is less than $1.3 + 0.06(G_{\rm BP}-G_{\rm RP})^2$;
  \item The Gaia XP spectra signal-to-noise ratio (SNR) for the full wavelength range is greater than 15;
  \item The extinction errors $E_{\rm err}$ from \citet{Zhang2023} are less than 0.02\,mag.
\end{enumerate}

Following these criteria, we have obtained a total of 2,974,582 stars for our Calibration Sample. From this sample, we further select a `Control Sample' with relatively small dust extinction values. This sample is confined to a specific sky region defined by the following coordinates: right ascension $105 <  $RA$ < 110.0$\,deg, declination $-3.5 < $Dec$ < 3$\,deg, and Galactic latitude $b > 3.5$\,deg. The Control Sample comprises 18,002 stars. Fig.~\ref{spadist} illustrates the spatial distribution of stars in both the Calibration and Control Samples, as well as the total line-of-sight (LOS) extinction for these regions. Thanks to the extensive stellar count and all-sky distribution of stars provided by the Gaia XP spectra catalogue, we have obtained a substantial VPHAS+ $u$-band Calibration Sample. This sample is distributed across each $u$-band observation visit, a feat unachievable with traditional SCR methods which adopt spectroscopic catalogues. The extinction values are derived from two-dimensional extinction maps by \citet{Planck2014}. The areas covered by the VPHAS+ survey generally exhibit significant LOS extinction. We selected areas with relatively low LOS extinction as reference regions to minimize the impact of extinction on the $u$-band photometric precision, thus obtaining a reference sample with more uniform photometry. The maximum extinction $E(B-V)$ in the Control Sample region does not exceed 0.4\,mag.

\begin{figure}
\centering
	\includegraphics[width=0.98\columnwidth]{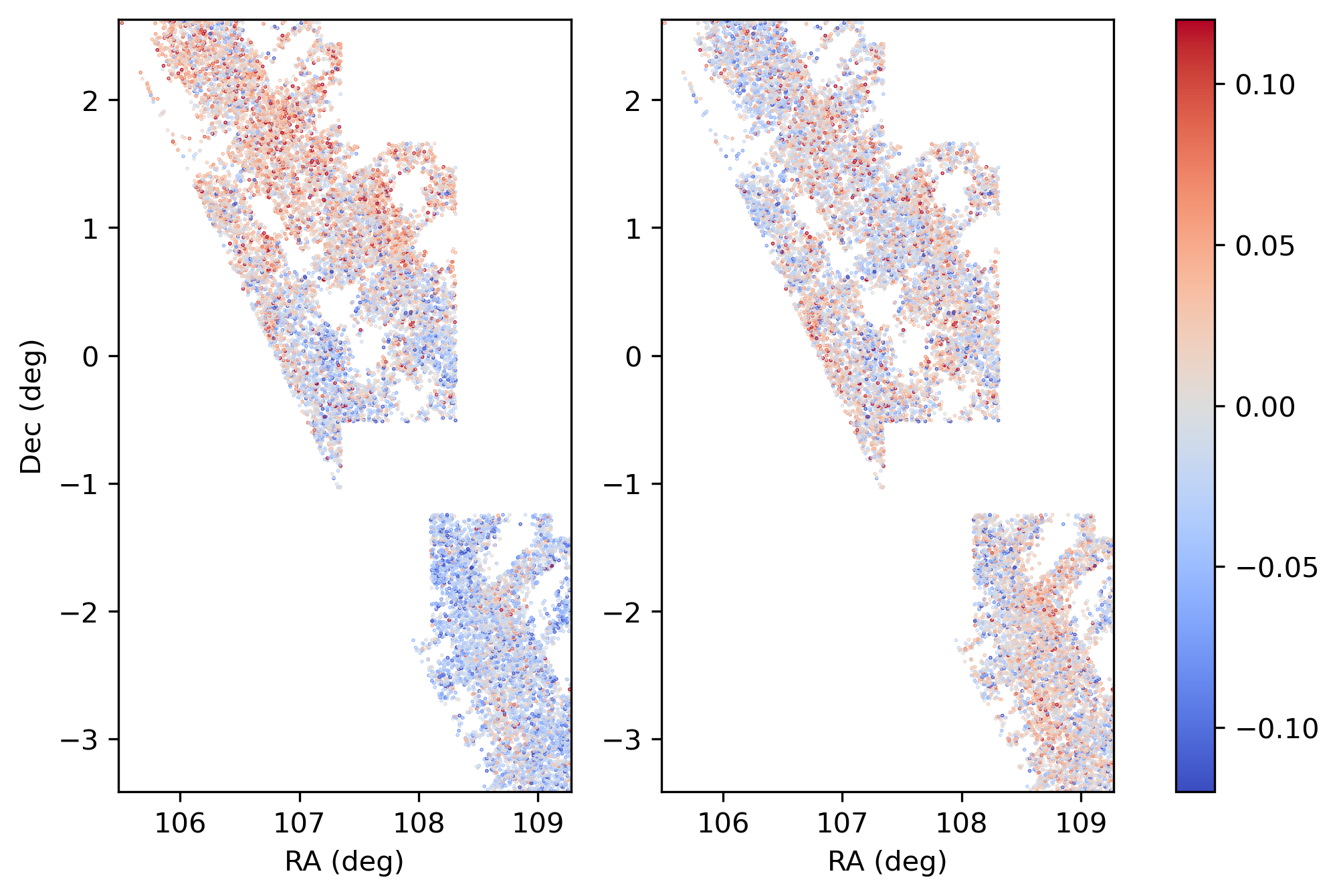}
    \caption{Spatial distribution of the $u$-band magnitude offset $\Delta u$ for the Control Sample before (left) and after (right) the preliminary photometric calibration. The colour scales represent the magnitude offsets for individual stars.}
    \label{ctsamp}
\end{figure}

\section{Method}
In this work, we adopt an updated SCR \citep{Yuan2015} method. Unlike the initial implementation of SCR, which predicted intrinsic stellar colours based on stellar atmospheric parameters such as effective temperature $T_{\rm eff}$ and metallicity [Fe/H] from spectroscopic surveys, our method predicts intrinsic stellar colours from intrinsic Gaia XP spectra. The detailed methodology employed is as follows.

\begin{enumerate}
    \item We begin by establishing the relationship between a star's intrinsic XP spectrum and its intrinsic colour $(u-G_{\rm BP})_0$ using the selected Control Sample. When obtaining the intrinsic XP spectra, we have applied a fixed extinction coefficient at each wavelength, sourced from \citet{Zhang2023}. This approach is justified given that the wavelength range corresponding to each flux measurement is only 10\,nm. However, to obtain the intrinsic colour $(u-G_{\rm BP})_0$, considering the broad $G_{\rm BP}$ band and the $u$ band that covers the near-ultraviolet portion of the stellar SED where variations are prominent, the extinction coefficient $R = E(u - G_{\rm BP})/E0$ cannot be assumed as a constant (e.g., \citealt{ZhangYuan2023, Sun2023}). Here, $E0$ is the extinction from \citet{Zhang2023}, and $R$ must at least vary with the effective temperature (or intrinsic colour) of the star. Initially, we use the extinction coefficient $R$ values provided by \citet{Zhang2023} as a starting point for calculating the intrinsic colour $(u-G_{\rm BP})_0$.
    
    After establishing the intrinsic colour $(u-G_{\rm BP})_0$, we employ a machine learning method, random forest regression (RFR), to train the relationship between the intrinsic XP spectra and $(u-G_{\rm BP})_0$. We randomly divide the Control Sample, using 75\% for training and 25\% for testing. The RFR model's input parameters are the standardized intrinsic XP spectra ($F' = (F-\mu)/\sigma$, where $F$ and $F'$ are the original and standardized fluxes, respectively, and $\mu$ and $\sigma$ are the mean and standard deviation). The output parameter is $(u-G_{\rm BP})_0$. This study uses the Python SCIKIT-LEARN package \citep{Pedregosa2011} to train the RFR model. After testing a range of RFR model parameters, we select the best-performing set for training.
    
    \item The RFR model obtained from the previous step will be applied to the intrinsic XP spectra of stars in the Calibration Sample to obtain the predicted colour $(u-G_{\rm BP})^{\rm mod}_0$. Thus, we can determine the reddening value for each star in the Calibration Sample as $E(u - G_{\rm BP}) = (u-G_{\rm BP}) - (u-G_{\rm BP})^{\rm mod}_0$. By combining this with the extinction values $E0$ provided by \citet{Zhang2023}, we can fit the extinction coefficient $R$ and the zero-point offset $\Delta M$. We use the following formula to relate $E(u - G_{\rm BP})$ to $E0$:
    \begin{equation}
    E(u - G_{\rm BP}) = (a_0C_0^2 + a_1C_0 + a_2)E0 + \delta M,
    \end{equation}
    where $C_0$ is the intrinsic colour $(u-G_{\rm BP})^{\rm mod}_0$, $a_0$, $a_1$, and $a_2$ are coefficients describing how the extinction coefficient $R$ varies with intrinsic colour $C_0$, and $\delta M$ represents the zero-point differences between the control and calibration samples. We iterate this step with the newly obtained extinction coefficients fed back into step~(i), limiting the iterations to no more than five, as results generally stabilize after two or three iterations.
    
    \item With the relationship between intrinsic stellar spectra and colours obtained in step (i), and the extinction coefficient $R$ obtained in step~(ii), we can predict the model magnitudes for each star in the Calibration Sample:
    \begin{equation}
    u_{\rm mod} = G_{\rm BP} + (u-G_{\rm BP})^{\rm mod}_0 + R \times E0.
    \end{equation}
    The magnitude offsets $\Delta u$ between the model and observed magnitudes can then be determined using:
    \begin{equation}
    \Delta u = u_{\rm mod} - u_{\rm obs} - \delta M.
    \end{equation}
    After obtaining the initial magnitude offsets $\Delta u$, we have performed a preliminary photometric calibration on the Control Sample to achieve a more uniform photometry across the reference stars. This preliminary calibration entails a straightforward adjustment of the zero-point differences in the magnitudes observed during each visit. We have computed the average $\Delta u$ for the Calibration Sample stars across each observational visit and applied the correction to derive the calibrated $u$-band magnitudes: $u_1 = u + \Delta u(\text{visit})$, where $u_1$ and $u$ are the post- and pre-calibration $u$-band magnitudes respectively, and $\Delta u(\text{visit})$ represents the average magnitude offset for each observational visit. The corrected magnitudes are then fed back into step~(i) to iterate the calibration process.

\end{enumerate}

Fig.~\ref{ctsamp} illustrates the spatial distribution of the $\Delta u$ for the Control Sample both before and after the preliminary flux calibration following the first iteration. Prior to the correction, a clear spatial variation in the magnitude offsets can be observed, with larger offsets in the upper-left region of the Diagram (Dec $> 0$\,deg) compared to the lower-right region (Dec $< -1$\,deg). The post-correction panel shows a more uniform distribution of magnitude offsets across different visits.

\begin{figure}
\centering
	\includegraphics[width=0.98\columnwidth]{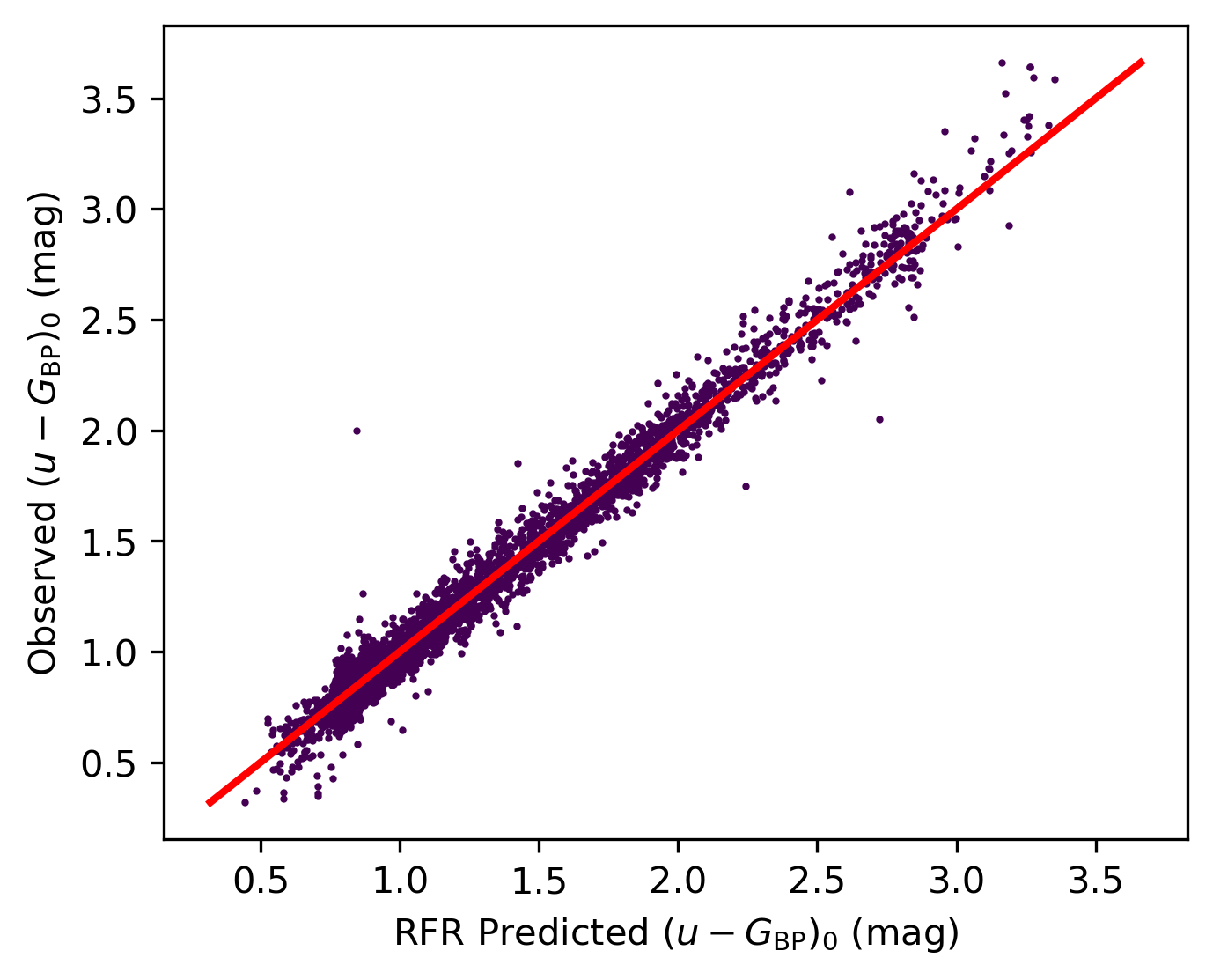}
    \caption{Comparison of intrinsic colour $(u-G_{\rm BP})_0$ between the observed values and those predicted by our RFR model for the test sample stars. The red line representing perfect agreement is plotted to guide the eye.}
    \label{fitcolor}
\end{figure}

\begin{figure}
\centering
	\includegraphics[width=0.98\columnwidth]{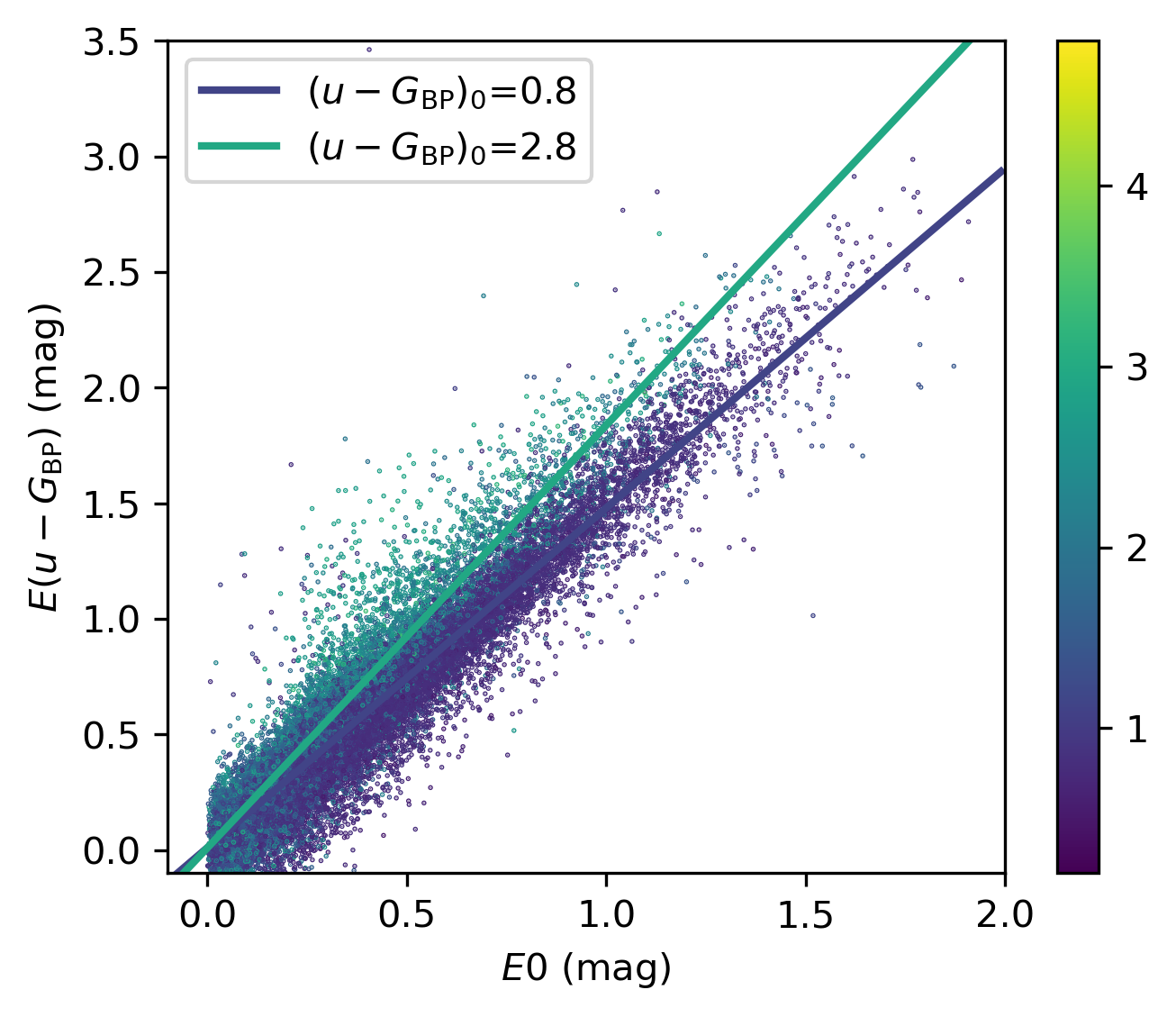}
    \caption{Linear regression of the reddening coefficients $E(u-G_{\rm BP})$ (this paper) with respect to $E_0$ \citep{Zhang2023} for the calibration sample. For clarity, only every 100th star is plotted. The colour scale indicates different intrinsic stellar colours $(u-G_{\rm BP})_0$. The fitted reddening coefficients for two representative intrinsic colours, $(u-G_{\rm BP})_0 = 0.8$ and $2.8$, are shown in the plot.}
    \label{fitrubp}
\end{figure}

\begin{figure*}
    \centering
    \includegraphics[width=0.98\textwidth]{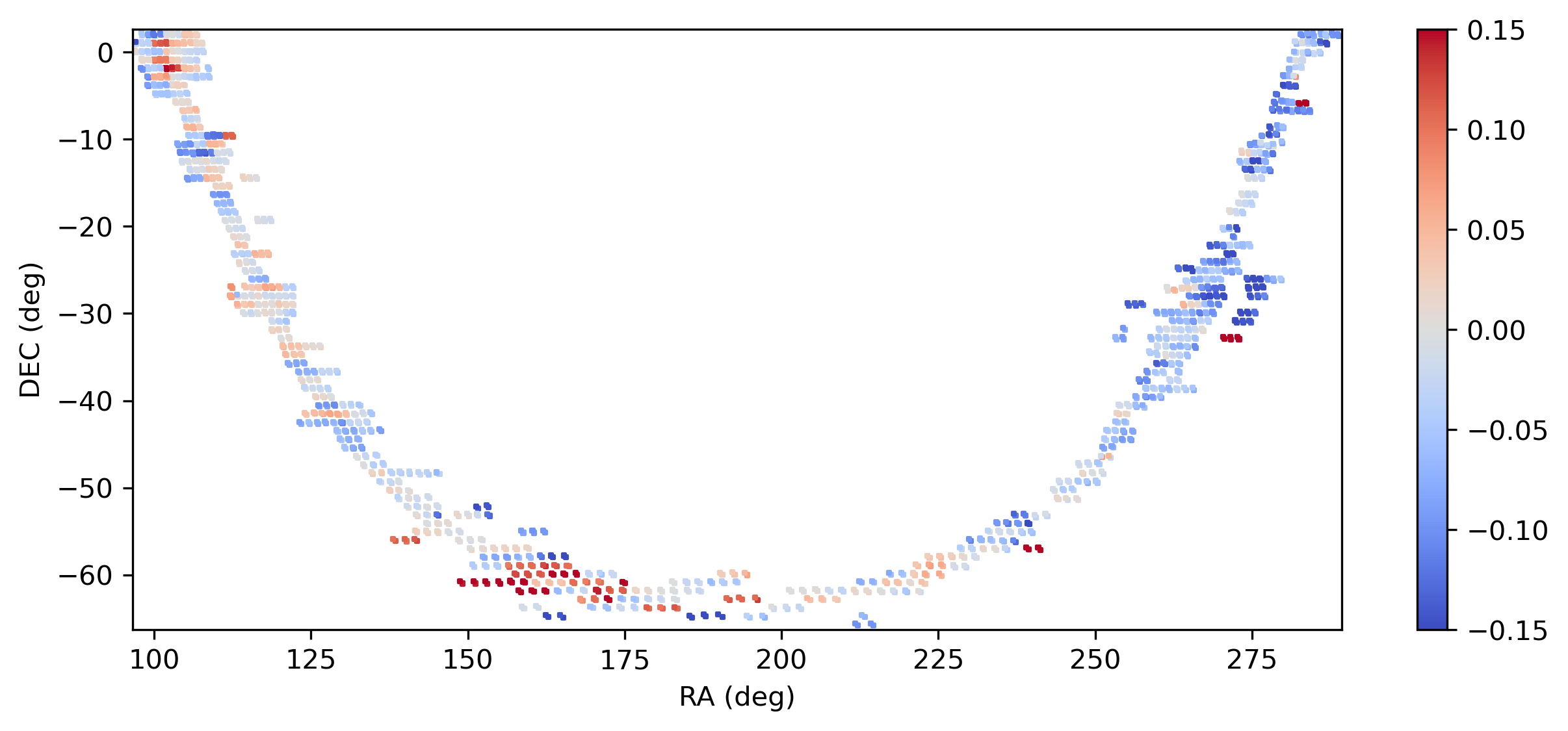}
    \caption{Spatial variation of the mean stellar magnitude difference for each visit, $\Delta u_1$(visit), in equatorial coordinates. The colour scale corresponds to the $\Delta u_1$(visit) values.}
    \label{fig:avg_delta_u_space}
\end{figure*}

\begin{figure*}
    \centering
    \includegraphics[width=0.98\textwidth]{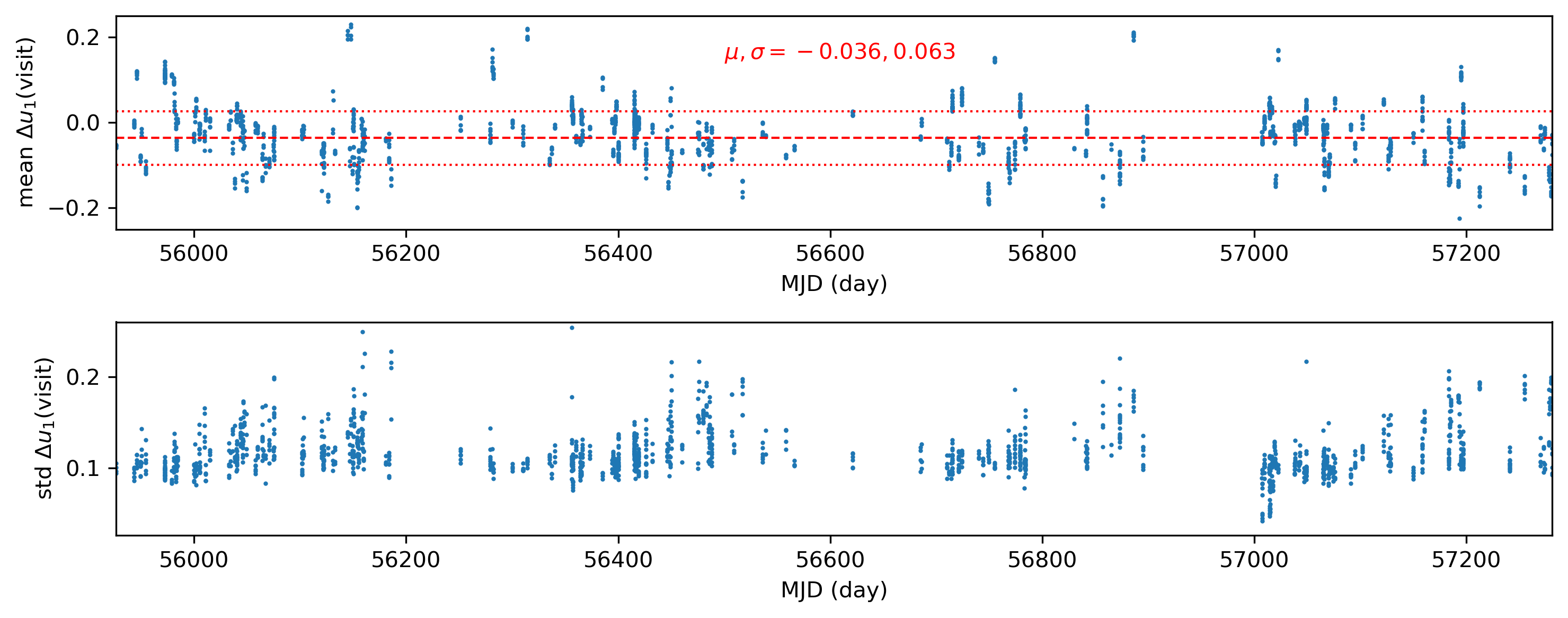}
    \caption{Temporal variation of the mean (upper panel) and standard deviation (lower panel) of the magnitude difference throughout each observational visit as a function of Modified Julian Date (MJD). The upper panel features a red dashed line indicating the mean, and a red dotted line representing the standard deviation of $\Delta u_1({\text{visit}})$, both of which are also annotated within the panel for clarity.}
    \label{fig:avg_delta_u_time}
\end{figure*}

\begin{figure}
\centering
    \includegraphics[width=0.48\columnwidth]{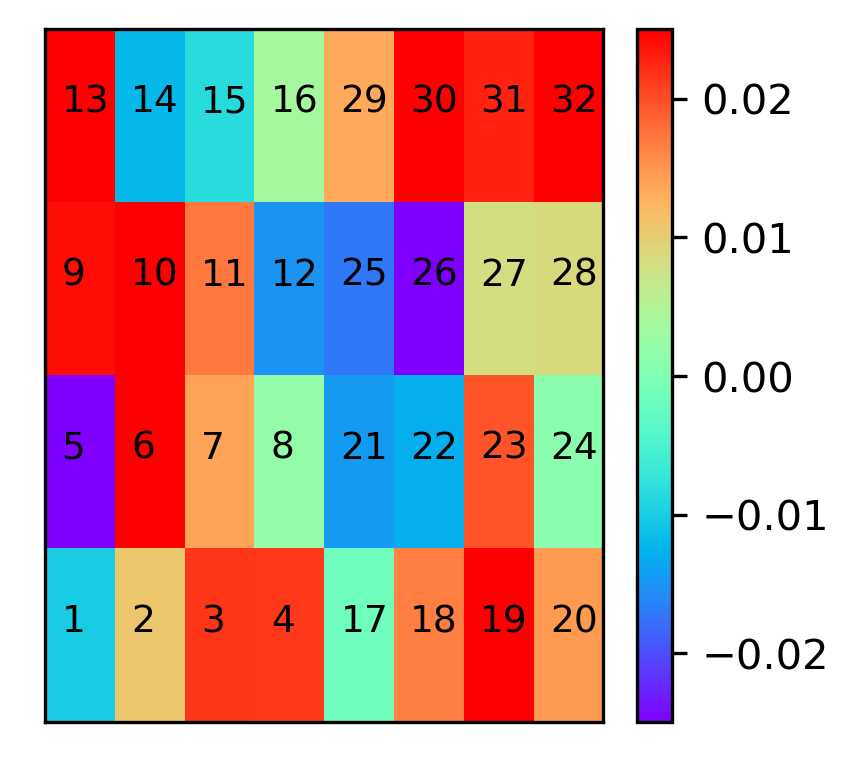}
    \includegraphics[width=0.48\columnwidth]{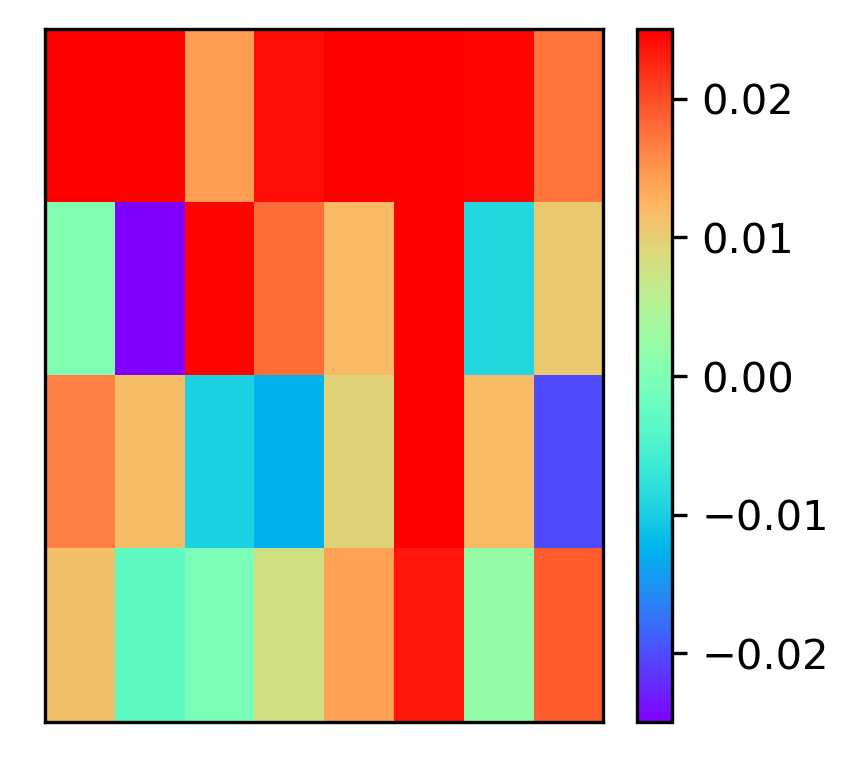}
    \caption{Distribution of the mean magnitude discrepancies $\Delta u_2$ across various CCDs for two distinct observation sessions (MJD 55926.213 and MJD 55943.313 are depicted in the left and right panels, respectively). The colour scale represents the value of $\Delta u_2$(CCD, visit). CCD identifiers are annotated in the left panel.}
    \label{ccdmean}
\end{figure}

\begin{figure*}
\centering
    \includegraphics[width=0.98\textwidth]{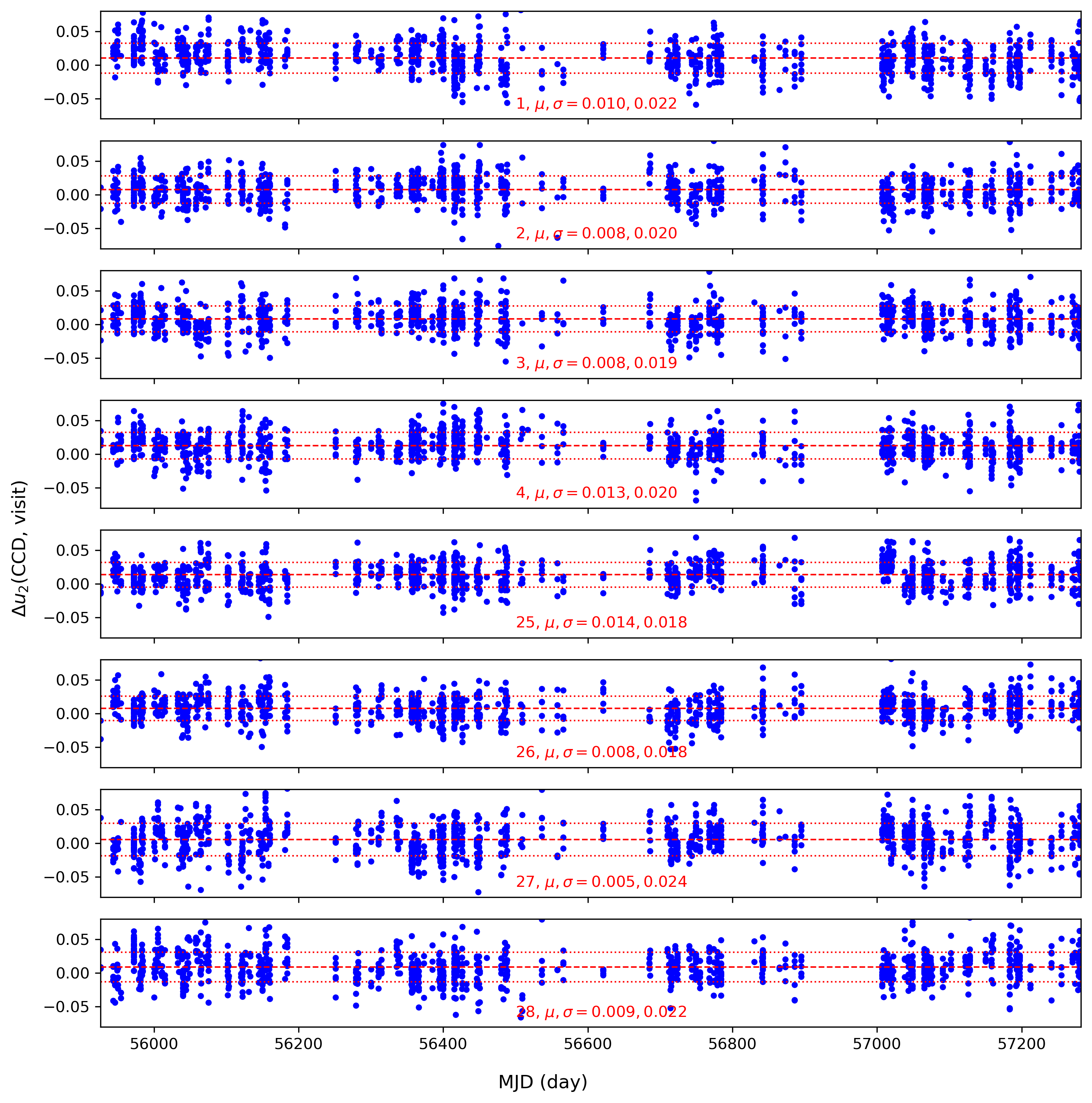}
    \caption{The temporal variation of $\Delta u_2$(CCD, visit) for a selection of eight CCDs, plotted as a function of MJD. Red dashed lines indicating the mean, and red dotted lines representing the standard deviation of $\Delta u_2({\text{CCD,~visit}})$ are overplotted in the diagrams. Annotations include CCD numbers (see Fig.~\ref{ccdmean}) and the corresponding mean and standard deviation of $\Delta u_2$(CCD, visit), highlighted in red for each graph.}
    \label{ccdmjd}
\end{figure*}

\begin{figure}
\centering
    \includegraphics[width=0.48\textwidth]{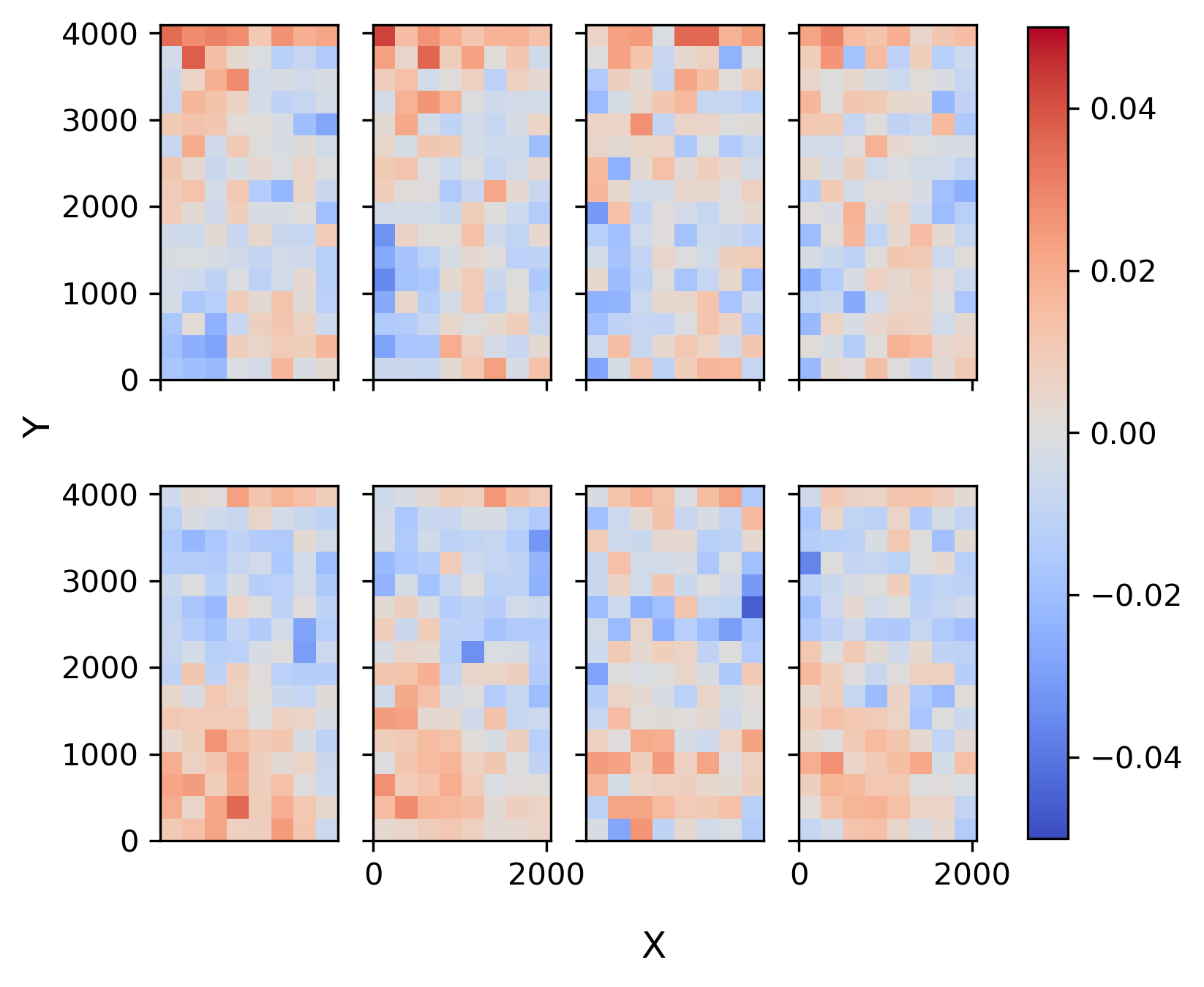}
    \caption{Spatial distribution of $\Delta u_3(\text{pixel})$ within the $X-Y$ plane for CCDs numbered 1 (top panels) and 25 (bottom panels) over four observational quarters spanning from October 2011 to October 2015. The colour scale denotes the mean $\Delta u_3$ within each $256 \times 256$ pixel bin. Sequential panels represent consecutive observing quarters from left to right.}
    \label{ccdvari}
\end{figure}

\begin{figure*}
\centering
    \includegraphics[width=0.88\textwidth]{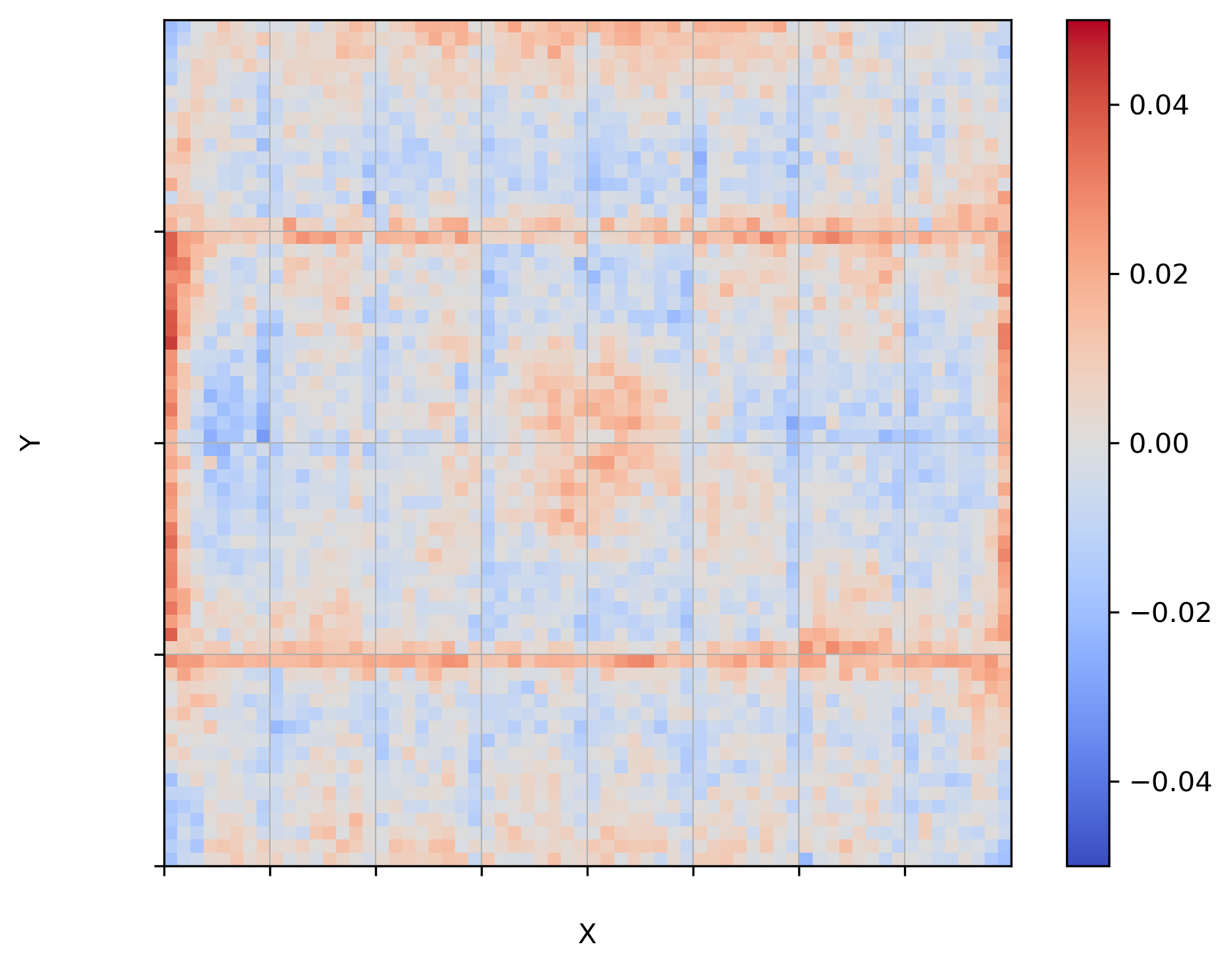}
    \caption{Variations of $\Delta u_3(\text{pixel})$ values within the entire CCD array in the $X-Y$ pixel space. The colour scale indicates the mean $\Delta u_3$ value for each $256 \times 256$ pixel bin.}
    \label{allccd}
\end{figure*}

\begin{figure*}
\centering
    \includegraphics[width=0.3\textwidth]{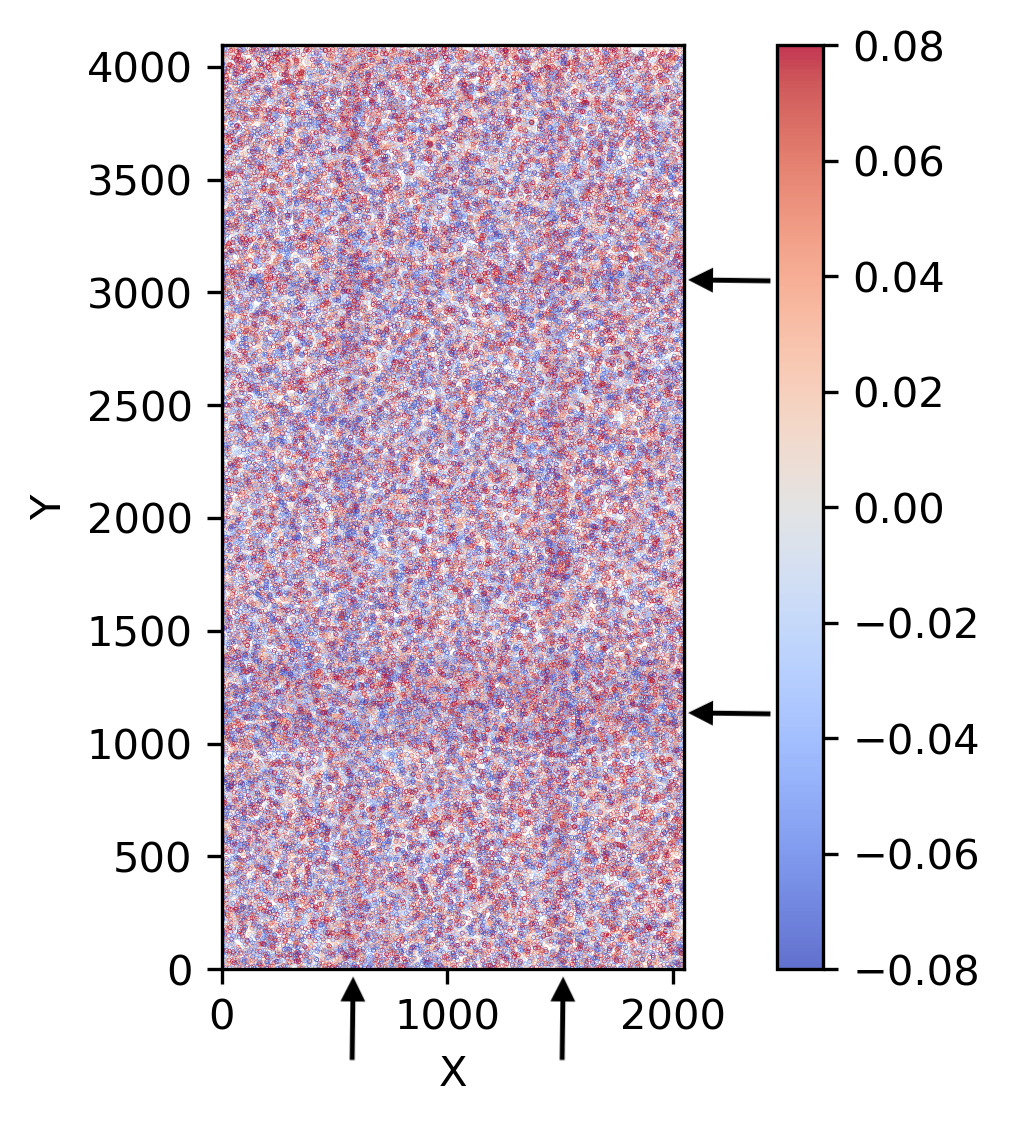}
    \includegraphics[width=0.3\textwidth]{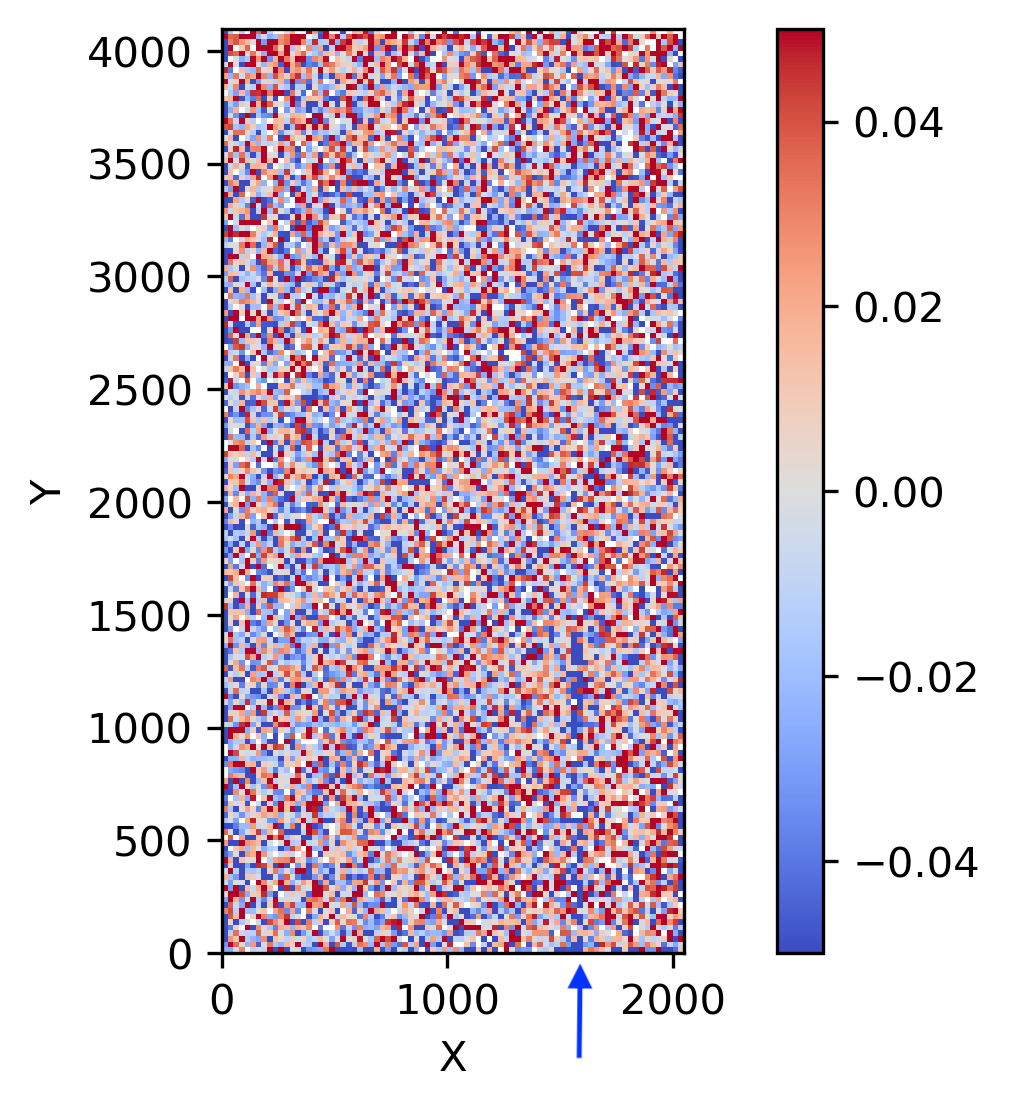}
    \includegraphics[width=0.3\textwidth]{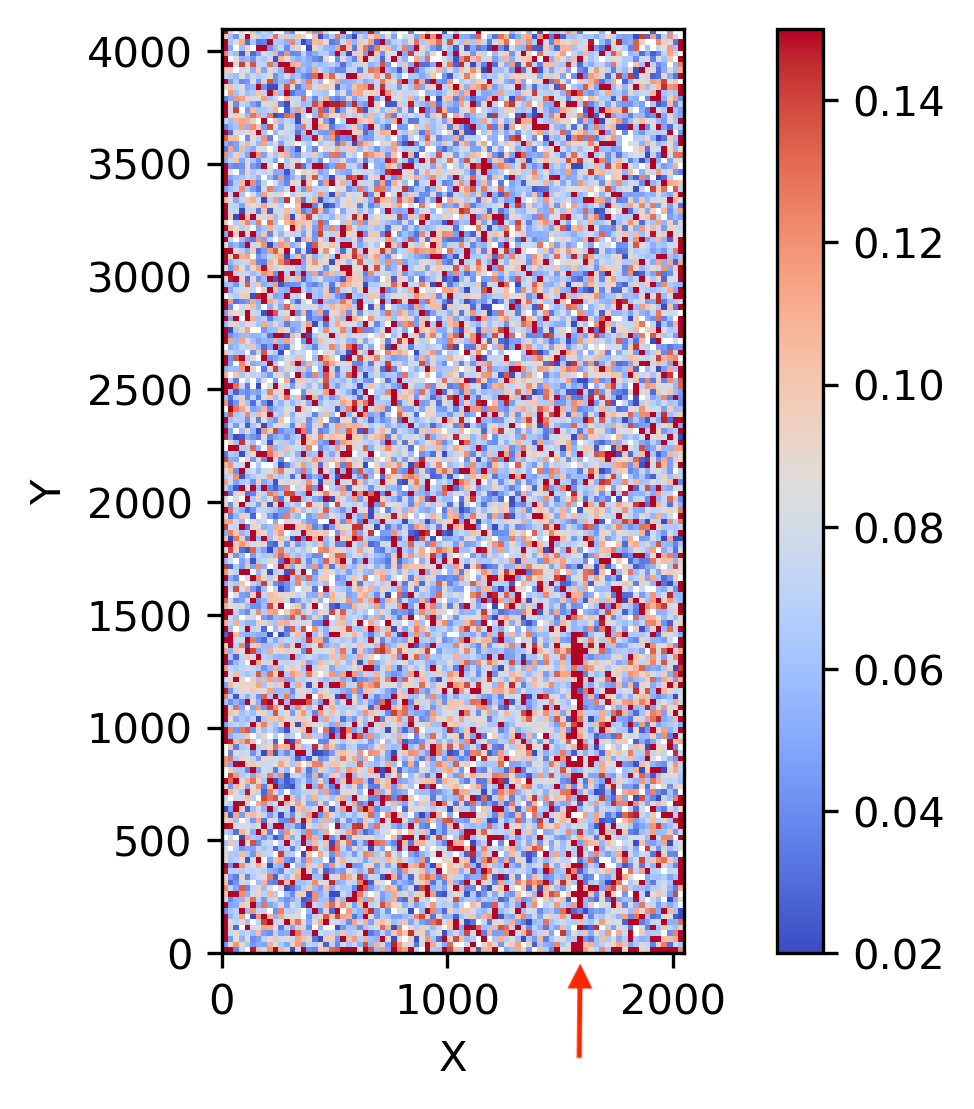}
    \includegraphics[width=0.3\textwidth]{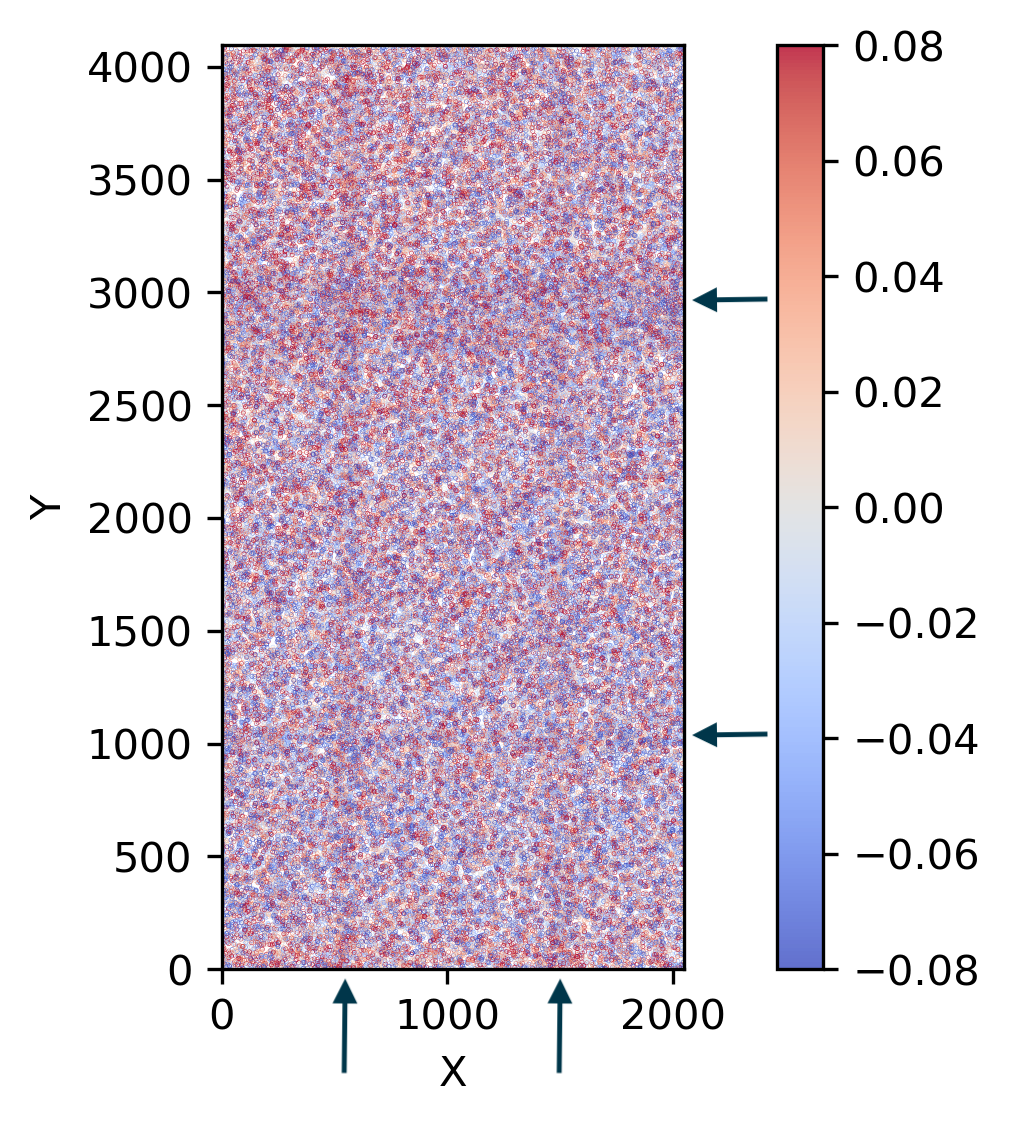}
    \includegraphics[width=0.3\textwidth]{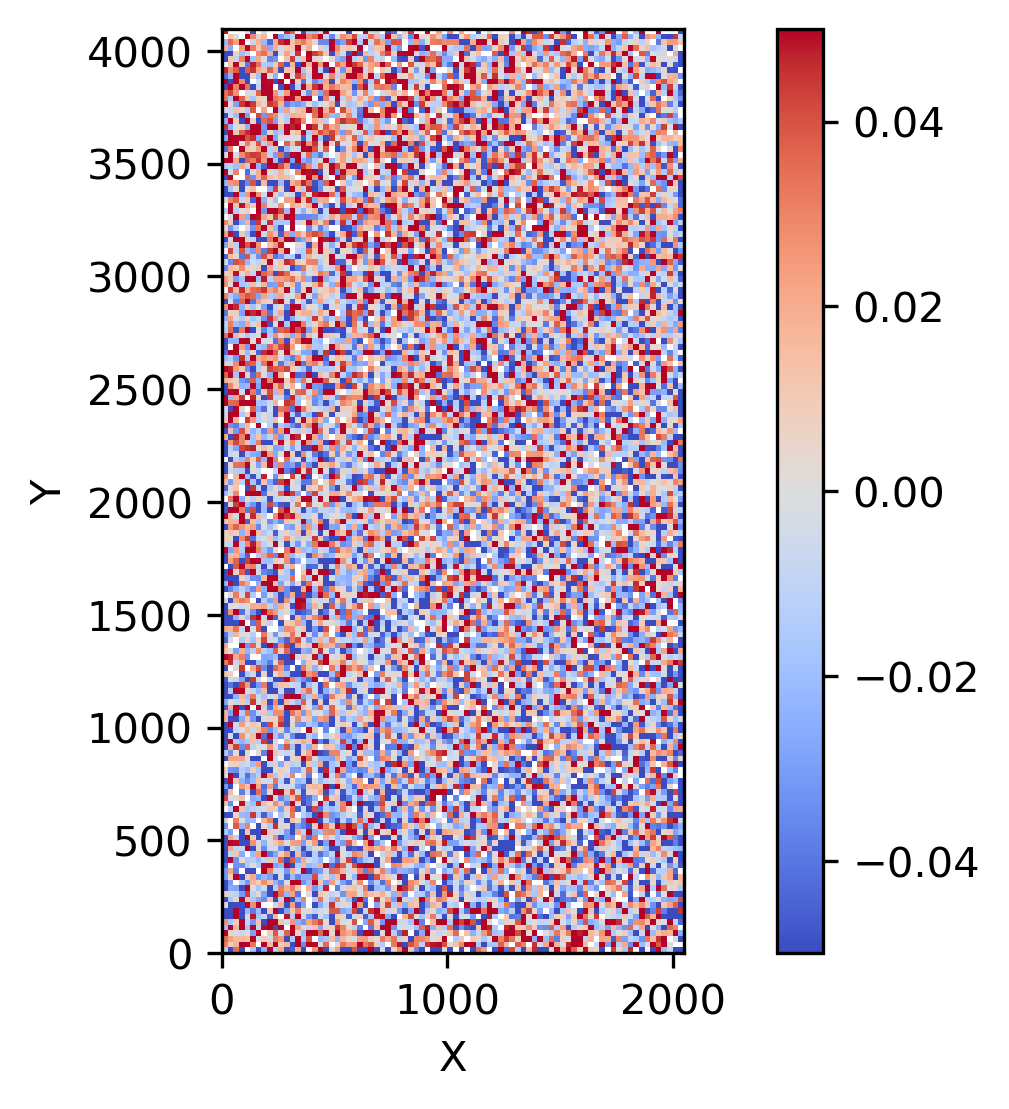}
    \includegraphics[width=0.3\textwidth]{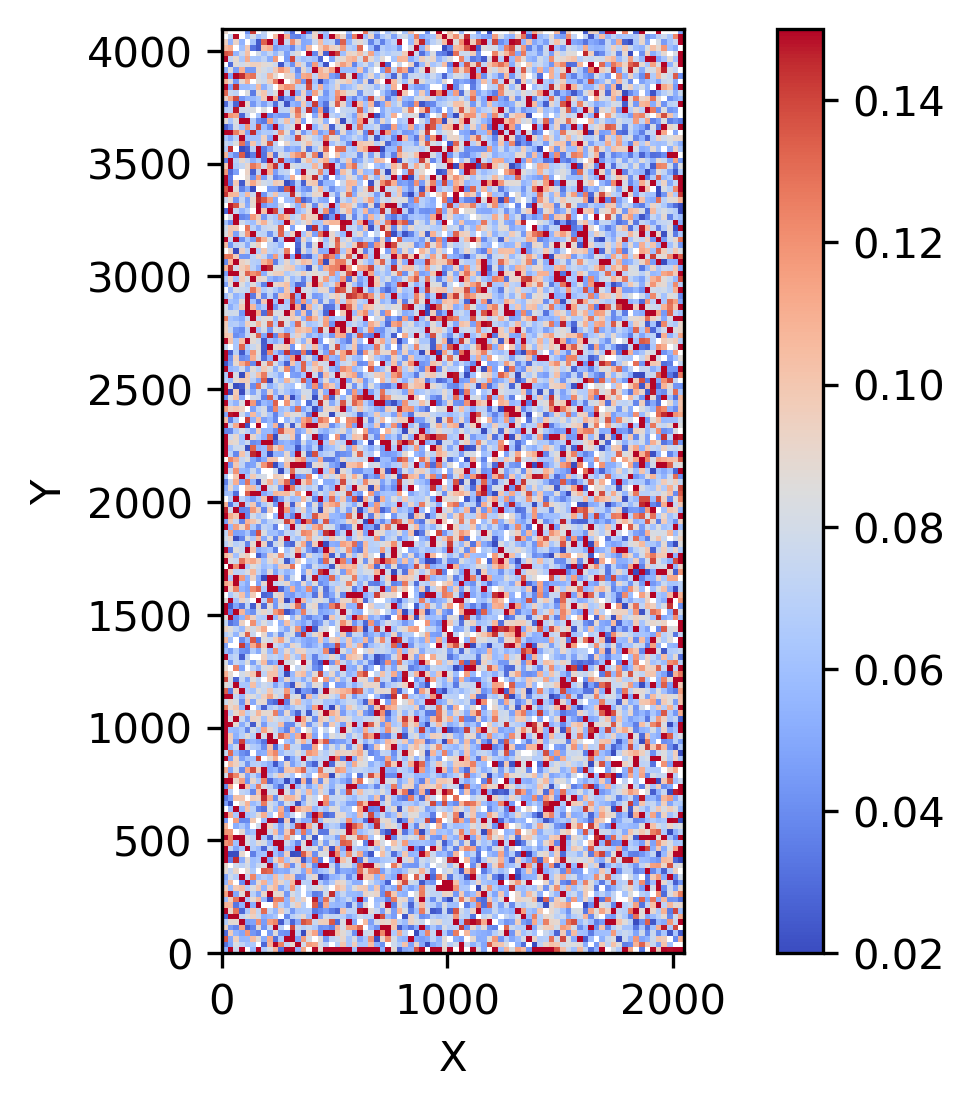}
    \caption{Variation of the single-star magnitude differences $\Delta u_3$ within CCD 11 (upper panels) and CCD 21 (lower panels). The $XY$ space distributions for individual stars are shown in the left panels. Mean magnitude differences and standard deviations for each $25 \times 25$ pixel bin are plotted in the middle and right panels, respectively. Black arrows in the left panels highlight the subtle cross features. A blue arrow in the upper middle panel and a red arrow in the upper right panel denote the location of a defective column.}
    \label{ccd1121}
\end{figure*}

\begin{figure}
\centering
\includegraphics[width=0.48\textwidth]{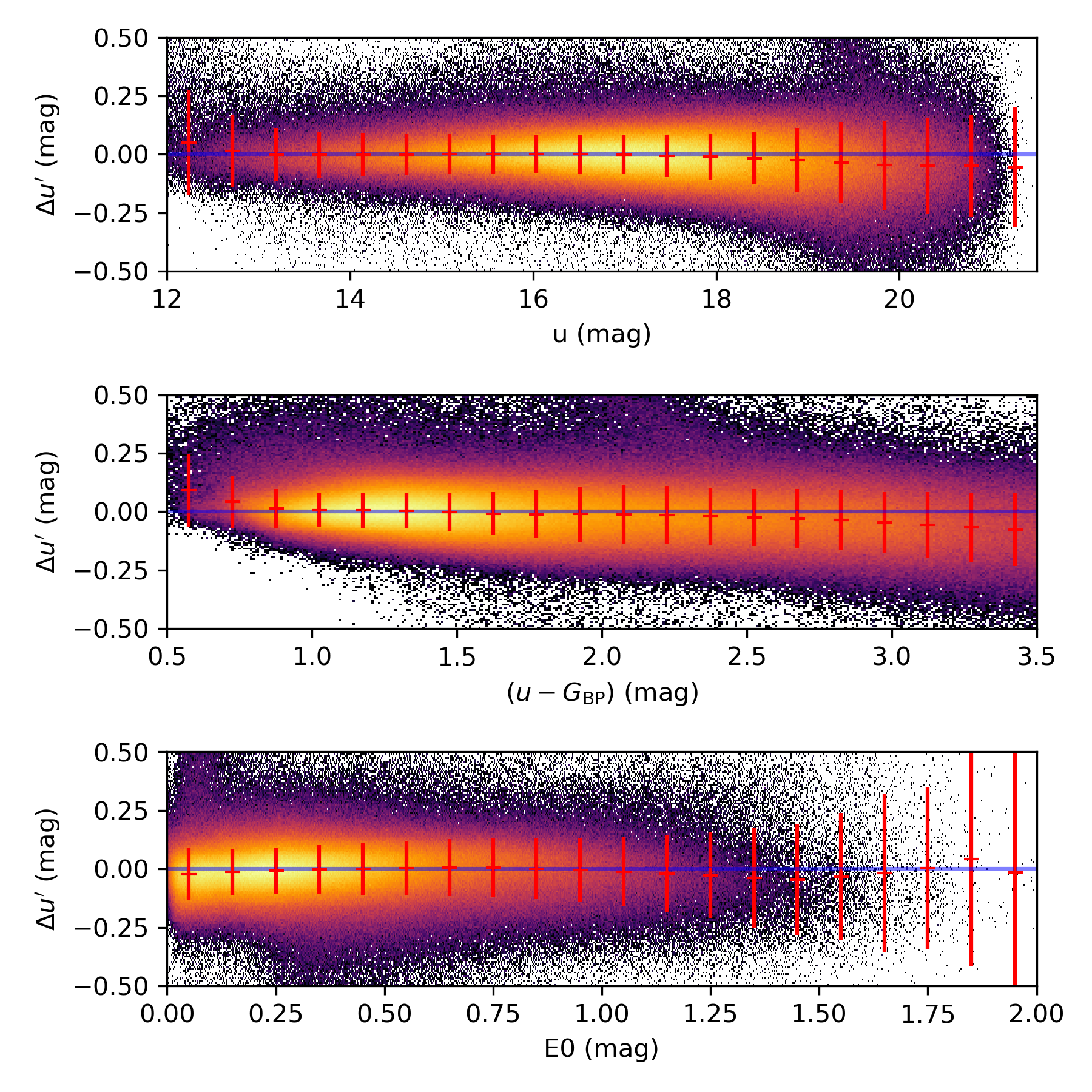}
\caption{The post-correction magnitude offset $\Delta u'$ plotted against stellar magnitude $u$ (upper panel), colour index $(u-G_{\text{BP}})$ (middle panel), and extinction $E0$ (lower panel). The colour scales visualize the density of stars in the Calibration Sample. Red points with error bars indicate the median and dispersion within different bins, and the solid blue lines represent the zero offset position where $\Delta u' = 0$. The photometric zero-point, $u_{\rm ZP}$, was determined using stars that fell within the magnitude range of $14 < u < 17$\,mag, the colour index range of $1 < (u-G_{\rm BP}) < 2$\,mag, and extinction values less than $E0 < 1.0$\,mag. }
\label{dumag}
\end{figure}

\begin{figure}
\centering
\includegraphics[width=0.48\textwidth]{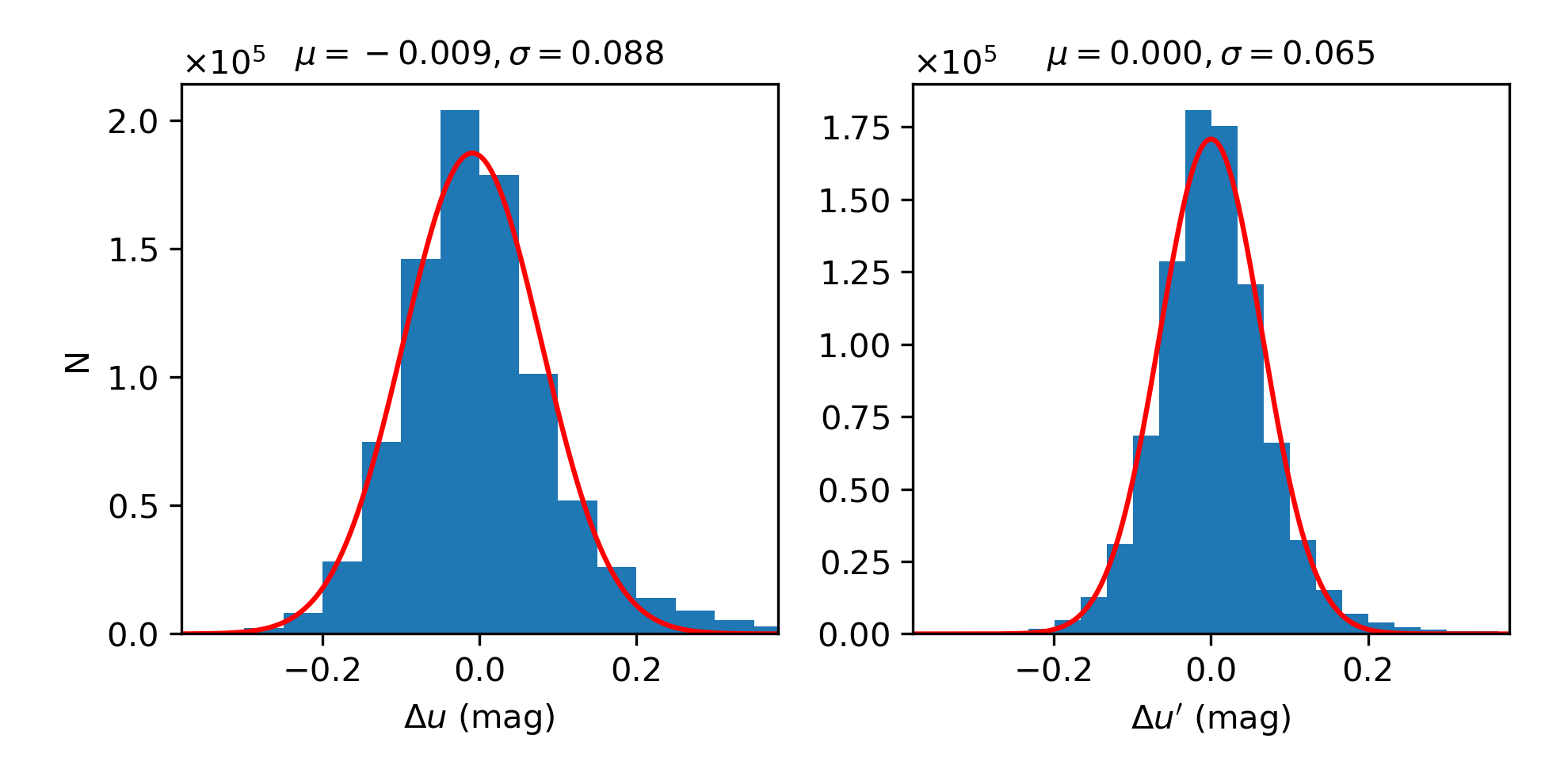}
\caption{Histogram distributions of magnitude offset before ($\Delta u$, left panel) and after ($\Delta u'$, right panel) correction. The red solid lines represent Gaussian fits, with the derived medians and dispersions annotated in each panel's title.}
\label{duhist}
\end{figure}

\begin{figure}
\centering
\includegraphics[width=0.48\textwidth]{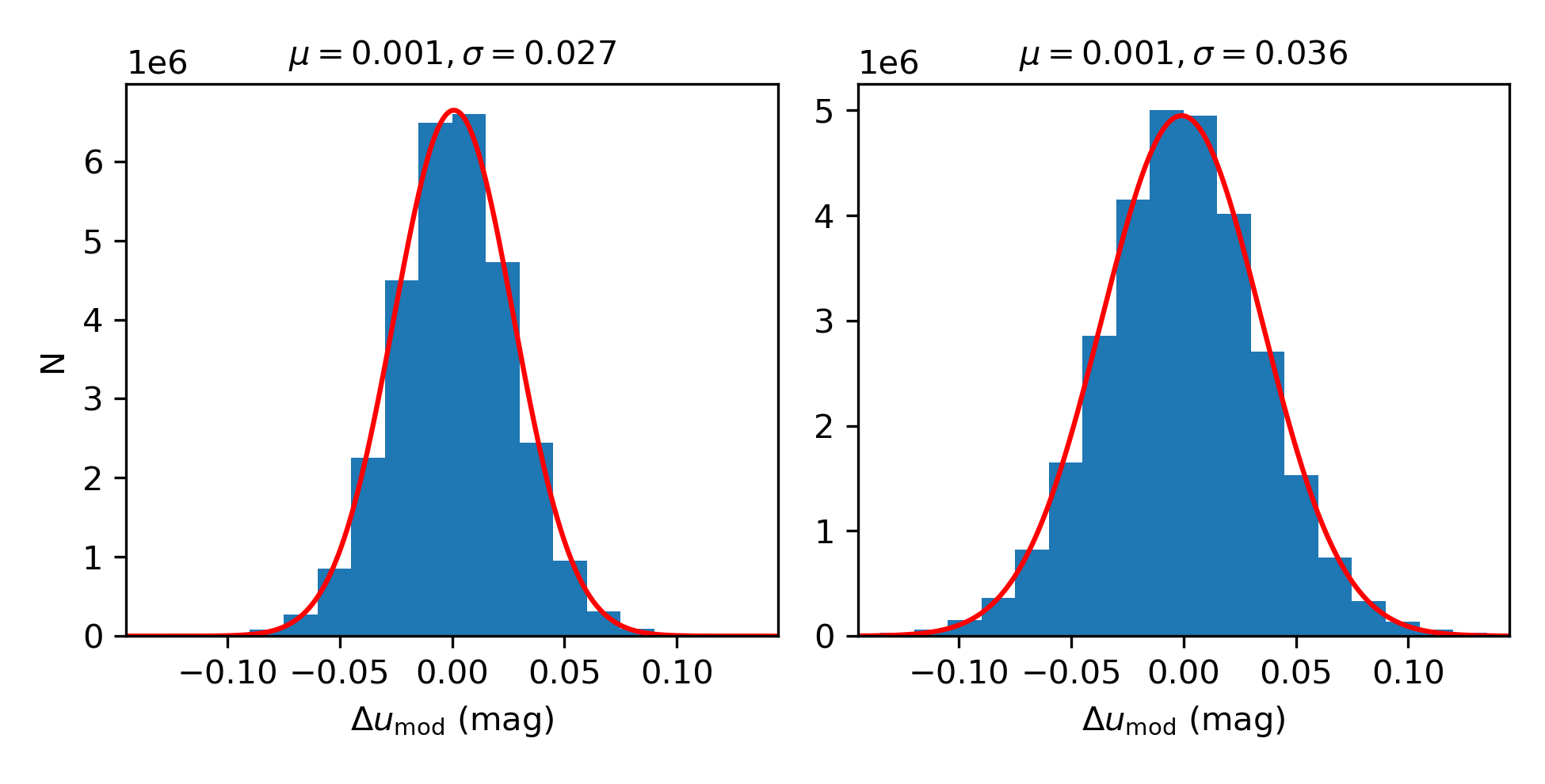}
\caption{Histograms of the differences $\Delta u_{\rm mod}$ between the model magnitudes from the Calibration Sample and those generated from the 10 MC simulations for the median set (left panel) and the 90th percentile set (right panel). The red curves represent Gaussian fits to the distributions, with the mean and standard deviation values provided in the titles of each panel.}
\label{mchist}
\end{figure}

\begin{figure}
\centering
\includegraphics[width=0.48\textwidth]{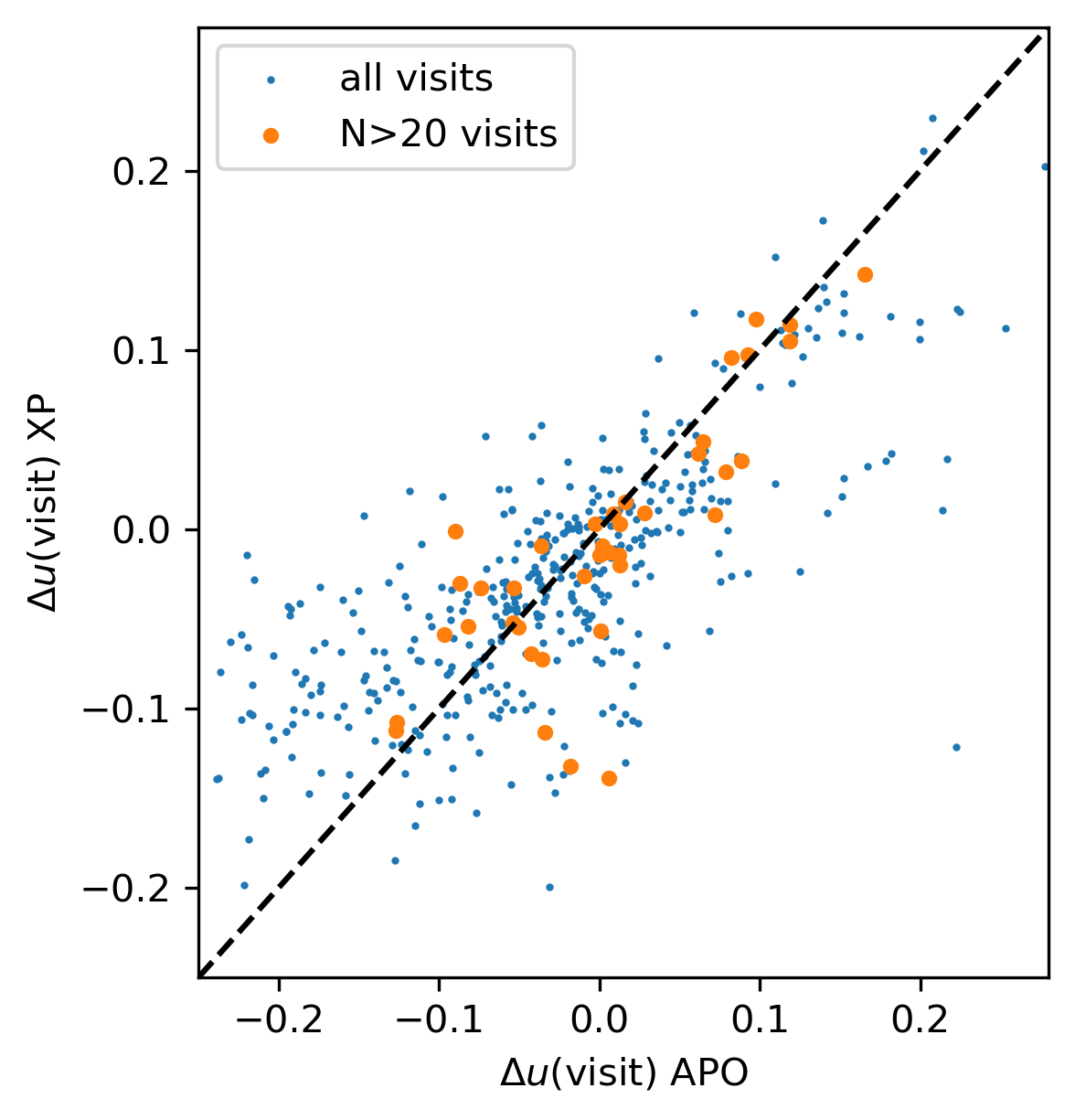}
\caption{Comparison of the mean magnitude offsets $\Delta u(\text{visit})$ for individual VPHAS+ observation visits calculated from APOGEE stellar parameters versus those derived from Gaia XP spectra. The blue dots represent all observation visits, while the red dots indicate visits with more than 20 stars calibrated using APOGEE data. The black dashed line denotes the line of perfect agreement and is included to facilitate visual comparison.}
\label{compapo}
\end{figure}

\section{Result and discussion}

We first present some intermediate results obtained during the process of deriving the $u$-band model magnitudes. For training, a subset consisting of 25\,per\,cent of the Control Sample is randomly selected as the test sample to gauge the performance of our RFR model. Fig.~\ref{fitcolor} displays a comparison between the intrinsic colours $(u-G_{\rm BP})_0$ predicted by our RFR model for the test sample stars and the observed values. The prediction closely matches the observations with a consistency parameter, the $r^2$ score, reaching 0.98. The dispersion between predicted and observed values is 0.064\,mag. This outcome aligns with our expectations since the intrinsic XP spectra of stars contain essential information about their atmospheric parameters and intrinsic colours. In modelling the intrinsic colours, we utilized spectral fluxes directly, avoiding possible systematic errors inherent in measuring atmospheric parameters \citep{Shen2022, Yang2024}. 

Fig.~\ref{fitrubp} shows the relationship between the reddening values $E(u-G_{\rm BP})$ for the Calibration Sample stars and the $E0$ values from \citet{Zhang2023}. There are clear differences in the $R$ values for stars of different intrinsic colours (or effective temperatures). Our fitted results, based on Equation~(1), correspond well with the data. The fitted relation for the extinction coefficient is as follows:
\begin{equation}
R = 0.042(u-G_{\rm BP})^2_0 + 0.027(u-G_{\rm BP})_0 + 1.42.
\end{equation}
The magnitude difference $\delta M$ between the Control Sample and the Calibration Sample is determined to be 0.009\,mag.

\subsection{Validating the $u$-band photometric calibration}

\subsubsection{Magnitude differences across observational visits}

Upon calculating the magnitude discrepancy, $\Delta u$, for each star within our Calibration Sample, we proceed to investigate the consistency of this discrepancy across separate observational visits. Fig.~\ref{fig:avg_delta_u_space} illustrates the spatial variability of the average $\Delta u$ values (denoted as $\Delta u_1({\text{visit}})$) relative to equatorial coordinates for each visit. A clear large-scale spatial variation in magnitude differences is evident, where $\Delta u_1$(visit) values tend to be similar within adjacent sky regions at the same declination. However, there is a more significant variation across different right ascension intervals. For instance, $\Delta u$(visit) is notably smaller in regions with RA $>$ 250\,deg, and significantly larger within the 150 $<$ RA $<$ 175\,deg interval.

We have also examined the temporal evolution of $\Delta u_1$(visit), as depicted in the upper panel of Fig.~\ref{fig:avg_delta_u_time}. The mean magnitude difference exhibits a distinct seasonal trend. This trend is more pronounced in the bottom panel of Fig.~\ref{fig:avg_delta_u_time}, which shows the dispersion of magnitude differences over time for each observational visit. Typically, starting from October of each observing season, which corresponds to the Southern Hemisphere's spring, the mean value of $\Delta u_1$(visit) approaches zero with a smaller dispersion, approximately 0.1\,mag. As time progresses, the mean increasingly deviates from zero, and the dispersion grows, reaching its maximum during the summer. These variations in $\Delta u_1$(visit) with time and space reflect the unaccounted observational conditions, such as seeing and atmospheric quality, during photometric measurements. Quantitatively, the mean and standard deviation of $\Delta u_1({\text{visit}})$ are $-0.036$\,mag and $0.063$\,mag, respectively.

\subsubsection{Magnitude differences across individual CCD chips}

We have then corrected the observed magnitude differences $\Delta u_1$(visit) for each target field, obtaining a new set of corrected magnitudes $u_2 = u_{\rm obs} + \Delta u_1$(\text{visit}). These newly derived magnitudes $u_2$ should exhibit greater uniformity across both large spatial scales and over time, though discrepancies may persist among different CCDs. To test for systematic photometric calibration differences across the CCD array, we have compiled the CCD identifiers of each Calibration Sample star as well as their $X$ and $Y$ image coordinates. We then examine the mean stellar magnitude difference $\Delta u_2 = u_{\rm mod} - u_2$ averaged across different CCDs for each visit. 

Fig.~\ref{ccdmean} displays the distribution of these mean magnitude discrepancy values, $\Delta u_2\mathrm{(CCD, vist)}$, for two arbitrarily selected observation sessions across different CCDs within the field of view (FOV). Post-correction for $\Delta u_1$(visit) on a large spatiotemporal scale, the new $\Delta u_2$(CCD, visit) values show reduced variation across different CCDs, generally lying within the range of $-0.02$ to $0.02$ mag. However, these variations also exhibit temporal fluctuations. 

Fig.~\ref{ccdmjd} illustrates the time evolution of the mean magnitude differences $\Delta u_2$(CCD, visit) for eight selected CCDs. CCDs numbered 1 through 4 reside at the FOV periphery (positioned in the lower-left quarter as depicted in Fig.~\ref{ccdmean}), while CCDs 25 through 28 are situated nearer to the FOV center. These figures collectively unmask a gradual, long-term shift in the mean magnitude discrepancy, $\Delta u_2$(CCD, visit).  This could be attributed to variations in the gain factors of the individual CCDs, as well as to imperfections in the flat-field correction process. The computed mean and standard deviation of $\Delta u_2$(CCD, visit) across the entire ensemble of CCDs for all observational visits are $0.0096$\,mag and $0.022$\,mag, respectively. 

\subsubsection{Magnitude differences across individual pixel bins}

The performance and accuracy of CCDs in astronomical observations can be influenced by a multitude of factors. These include, but are not limited to, electronic bias, defective pixels, pixel response non-uniformity, electronic interference, and variations in the optical throughput of the telescope system. Moreover, despite careful dark current subtraction and flat-field correction, some residual noise and non-uniform response may persist. This can introduce intermediate-scale flux variations in the astronomical data, potentially impacting the scientific interpretation of the measurements.

Corrections to the observed magnitude differences $\Delta u_2({\text{CCD,~visit}})$ are also applied for each CCD chip to derive new corrected magnitudes $u_3 = u_2 + \Delta u_2({\text{CCD,~visit}})$. We compile observational data for each observing quarter, which corresponds to a one-year span, and scrutinize the new magnitude differences $\Delta u_3 = u_{\mathrm{mod}} - u_3$ for stars in the Calibration Sample across individual pixels within each CCD. To reduce the influence of stochastic errors, we group stars within each $256 \times 256$ pixel area and calculate the mean magnitude offset, $\Delta u_3(\text{pixel})$.

Fig.~\ref{ccdvari} illustrates the spatial distribution of $\Delta u_3(\text{pixel})$ values across the $X-Y$ pixel coordinates for CCDs numbered 1 and 25, encompassing four years of observational quarters. Within an individual CCD, the average $\Delta u_3$ uncovers small-scale patterns—evident as larger values at the top of CCD 1 and smaller ones at the bottom, with an opposite pattern for CCD 25. Remarkably, these structural features persist with consistency through various observing quarters, suggesting intrinsic, temporally stable attributes.

In Fig.~\ref{allccd}, we amalgamate the data from all observing quarters and visualize the variation of the mean magnitude difference, $\Delta u_3(\text{pixel})$, within each CCD using the same $256 \times 256$ pixel binning scheme. The resulting patterns of magnitude offset variation display pronounced $\Delta u_3(\text{pixel})$ values at the periphery of the FOV and at the centre of the entire field. The composite mean and standard deviation of $\Delta u_3(\text{pixel})$ across all pixels are $0.0003$\,mag and $0.009$\,mag, respectively.

Finally, We turn our attention to the variability of individual star magnitude differences across the CCD array. Within a given CCD, a star's magnitude difference can fluctuate by as much as $\pm 0.1$\,mag. Moreover, nearly all CCDs exhibit faint structural features, characterized by two horizontal and two vertical bands, which are indicated by black arrows in Fig.~\ref{ccd1121}.

For a detailed analysis, we compute the mean magnitude differences and their standard deviations within each $25 \times 25$ pixel bin. The results suggest that these features become less pronounced, to the point of being nearly indiscernible, upon averaging. Stars aligned with these features tend to show marginally lower mean magnitude differences and slightly higher standard deviations.

In pursuit of the origins of these patterns, we have examined a selection of processed scientific images along with their respective twilight flat field images. However, these structural features are not apparent in the inspected images. Consequently, the origins of the faint horizontal and vertical bands remain elusive and are the subject of ongoing investigation.

Furthermore, we have identified a pronounced anomaly in the lower right corner of CCD 11, characterized by lower mean magnitude differences and increased standard deviations. It is depicted by a blue arrow in the mean value diagram and a red arrow in the standard deviation diagram in Fig.~\ref{ccd1121}. This aberration corresponds to a defective column on the CCD chip, a flaw that is readily apparent in the downloaded images.

\subsection{Enhancing the $u$-band photometric calibration}

In light of the findings presented earlier, we have refined the photometric calibration of the VPHAS+ $u$-band data. We introduce a correction that is dependent on the individual visits, CCDs, and pixels to enhance photometric accuracy. The corrected $u$-band magnitudes ($u_{\text{corr}}$) are calculated as follows:
\begin{equation}
u_{\text{corr}} = u_{\text{obs}} + \Delta u_1(\text{visit}) + \Delta u_2(\text{CCD,~visit}) + \Delta u_3(\text{pixel}) + u_{\text{ZP}},
\end{equation}
where $u_{\text{obs}}$ represents the observed magnitude, and $\Delta u_1(\text{visit})$, $\Delta u_2(\text{CCD, visit})$, and $\Delta u_3(\text{pixel})$ are the magnitude offsets averaged by the visits, CCDs, and pixel bins, respectively. These offsets are derived in the previous section. We calculate these offsets only when a minimum of 20 stars from the Calibration Sample is present within the visit, CCD, or pixel bin. We have not applied corrections for a subset of stars (1.6\,per\,cent of the Calibration Sample, amounting to 46,135 stars) due to the lack of sufficient star counts within their corresponding visit, CCD, or pixel bin. 

The term $u_{\text{ZP}}$ is akin to a flux calibration zero point. In this work, we have chosen to utilize model magnitudes as reference standards to compute $u_{\text{ZP}}$. This computation involves selecting stars that fall within the magnitude range of $14 < u < 17$\,mag, the colour index range of $1 < (u - G_{\text{BP}}) < 2$\,mag, and have extinction values less than $E0 < 1.0$\,mag. We then average the magnitude offsets $u_{\text{mod}} - (u_{\text{obs}} + \Delta u_1(\text{visit}) + \Delta u_2(\text{CCD, visit}) + \Delta u_3(\text{pixel}))$ for this subset of stars to determine the zero point $u_{\text{ZP}}$. Our findings yield a photometric zero point of $u_{\text{ZP}} = 0.009$\,mag.

Fig.~\ref{dumag} illustrates the post-correction magnitude offset, $\Delta u' = u_{\text{mod}} - u_{\text{corr}}$, as it relates to the stellar magnitude $u$, colour index $(u - G_{\text{BP}})$, and extinction $E0$. Consistent with our expectations, the post-correction magnitude offset generally approaches zero. Notable exceptions occur in regions with either a sparse stellar population or high extinction, which induce minor deviations from zero. We specifically select stars for which $\Delta u'$ is nearest to zero, confined to the magnitude range $14 < u < 17$\,mag, colour index range $1 < (u - G_{\text{BP}}) < 2$\,mag, and extinction $E0 < 1.0$\,mag. A Gaussian function is used to model the distribution of the magnitude offset both before ($\Delta u$) and after ($\Delta u'$) the correction is applied. The fitting results, shown in Fig.~\ref{duhist}, reveal that the pre-correction magnitude offset exhibits a systematic bias with a mean of $-0.009$\,mag and a standard deviation of 0.088\,mag. The post-correction analysis demonstrates the elimination of the systematic bias and a reduction in the standard deviation to 0.065\,mag. The improvement in photometric precision post-correction is approximately $\sqrt{0.088^2 - 0.065^2}$ = 0.059\,mag. This precision enhancement is consistent with the standard deviation of $\Delta u_1$(visit), which is 0.063\,mag (as shown in Fig.~\ref{fig:avg_delta_u_time}), which suggest that the systematic discrepancies in the VPHAS+ DR4 $u$-band data have been successfully mitigated.

\subsection{Estimation of Uncertainties in Model Magnitudes}

In this section, we analyze the magnitude errors of our new SCR method. The sources of error in model magnitudes primarily stem from uncertainties in observational data, the fitting errors of intrinsic colours using the RFR model, and the quadratic function fitting errors associated with the extinction coefficients. To quantify the uncertainties associated with the model magnitudes, we have adopted a Monte Carlo (MC) simulation approach. The observational inputs include photometric uncertainties from the VPHAS+ $u$ band and Gaia $G_{\rm BP}$ band, flux measurement errors in Gaia XP spectra, and the uncertainties in extinction $E0$ as reported by \citet{Zhang2023}. Our methodology involves defining two distinct sets of observational uncertainties: a `median set' representing the median errors across all data, and a `90th set' comprising errors at the 90th percentile, indicative of extreme error scenarios. 

For the median set, the errors in $u$, $G_{\rm BP}$, and $E0$ are 0.005\,mag, 0.003\,mag, and 0.014\,mag, respectively, while the Gaia XP spectra have an SNR of 118. For the 90th set, the errors in $u$, $G_{\rm BP}$, and $E0$ are 0.021\,mag, 0.005\,mag, and 0.019\,mag, respectively, and the Gaia XP spectra have an SNR of 57. For each uncertainty set, we generate 10 MC realizations, resulting in a total of 20 MC samples.

By applying our SCR algorithm to these 20 MC data sets, we derive 20 corresponding sets of revised model magnitudes. The variability among these new magnitudes, when compared to our initial Calibration Sample-derived values, provides a statistical measure of the model magnitude uncertainties. The distributions of the magnitude differences for both the median and 90th set MC samples are presented in Fig.~\ref{mchist}. Gaussian fits to these distributions indicate that the standard deviation is 0.027\,mag for the median set and 0.036\,mag for the 90th set.

It is important to note that in our MC analysis, we have not considered the potential impact of variations in the dust extinction law, specifically changes in the total-to-selective extinction coefficient $R_V$ \citep{ZYC2023}. Additionally, we acknowledge that the $u$-band magnitude errors reported in the VPHAS+ catalogue are significantly underestimated.

We have also calculated the $u$-band model magnitudes for stars in the VPHAS+ catalogue by employing atmospheric parameters provided by the Apache Point Observatory Galactic Evolution Experiment (APOGEE; \citealt{apogee}). These parameters include the effective temperature ($T_{\text{eff}}$) and metallicity ([M/H]), following a methodology similar to that described by \citet{Xiao2022}. Our analysis begins by selecting stars from the Sloan Digital Sky Survey Data Release 17 (SDSS DR17; \citealt{sdssdr17}) that have corresponding observations in APOGEE. These stars are then cross-matched with the VPHAS+ catalogue to form our primary dataset.

For the Calibration Sample, we have applied selection criteria that included: an SNR in APOGEE spectra greater than 20; $T_{\text{eff}}$ in the range of 4300--5300\,K; log\,$g$ values from 2.0 to 3.5\,dex; and [M/H] greater than $-$1\,dex. The Control Sample is further refined to include stars with right ascension less than 120\,deg, declination greater than $-7.5$\,deg, and reddening values $E(B-V)$ calculated from the APOGEE stellar parameters \citep{Chen2019b, Sun2023} less than 0.5\,mag. This selection process yields a total of 4,367 stars for the Calibration Sample and 532 stars for the Control Sample. We then determine the model magnitudes for the Calibration Sample using the same approach as \citet{Xiao2022}.

Owing to the limited sample sizes of Calibration and Control Sample stars from APOGEE, we refrain from a direct comparison of these model magnitudes with those derived from Gaia XP spectra. Instead, we first compare the APOGEE-based model magnitudes with the observed VPHAS+ magnitudes to derive a magnitude offset. We proceed by computing the average magnitude offset for each VPHAS+ observation visit, denoted as $\Delta u(\text{visit})$. This is subsequently compared to the average offset obtained from the Gaia XP spectra. The comparative analysis, depicted in Fig.~\ref{compapo}, demonstrates a strong agreement between the traditional SCR method utilizing APOGEE stellar parameters and the updated SCR method based on Gaia XP spectra. For observation visits with more than 20 stars, the standard deviation of the differences in $\Delta u(\text{visit})$ between the two approaches is 0.036\,mag.

\section{Conclusions}

Based on the corrected XP spectra \citep{Huang2024} and the $G_{\rm BP}$ photometric data from Gaia DR3, along with the individual stellar extinction values provided by \citet{Zhang2023}, we have applied the novel SCR method to valid and enhance the photometric calibration of the $u$-band magnitudes for the VPHAS+ DR4. In this work, we improve the previously established SCR approach by employing a machine learning technique, specifically RFR, to train the relationship between the extinction-corrected intrinsic XP spectra and the intrinsic stellar colours. Our method leverages the vast number of XP spectra available in Gaia DR3, effectively overcoming the limitations of traditional SCR methods, which rely heavily on spectroscopic survey data. As a result, we obtain a high-quality Calibration Sample of nearly 3 million stars, even for the surveys near the Galactic centre of VPHAS+. The RFR model, trained to predict the intrinsic colours of stars based on their XP spectra, shows remarkable consistency with the actual observations, achieving an $r^2$ score of 0.98. We also explore the relationship between colour excess $E(u-G{\rm BP})$ and extinction values $E0$ from \citet{Zhang2023}, observing how the extinction coefficients vary with intrinsic colour.

In this study, we have determined the model $u$-band magnitudes for stars in our Calibration Sample. Through our analysis, we have unveiled spatial and temporal variations in the $u$-band magnitude offsets within the VPHAS+ dataset, potentially linked to differing observational conditions. For each observational visit, the computed average magnitude offset, $\Delta u_1({\text{visit}})$, exhibits a mean value of $-0.036$\,mag and a standard deviation of $0.063$\,mag. Moreover, we have discerned variations in the magnitude offsets among different CCD chips within a single visit, which also fluctuate over time. These differences, denoted as $\Delta u_2(\text{CCD, visit})$, yield a mean value of $0.0096$\,mag and a standard deviation of $0.022$\,mag. Additionally, we report the detection of consistent magnitude differences at the pixel level within each CCD, which appear to be stable over time, with a mean offset $\Delta u_3{\text{(pixel)}}$ of $0.0003$\,mag and a standard deviation of $0.009$\,mag.

Upon correcting for these varying magnitude offsets across observational visits ($\Delta u_1({\text{visit}})$), CCD chips ($\Delta u_2(\text{CCD, visit})$), and pixel bins ($\Delta u_3{\text{(pixel)}}$), we have derived the revised $u$-band magnitudes. These corrections led to a reduction in the standard deviation between the corrected magnitudes and the model magnitudes, decreasing from $0.088$\,mag to $0.065$\,mag. Notably, the corrected magnitudes exhibit no residual dependence on the stellar magnitude, colour, or extinction values. This improvement underscores the significance of our corrections and the enhanced reliability of the $u$-band magnitude determinations as a result of this study.

\section*{Acknowledgements}

This work is partially supported by the National Key R\&D Program of China No. 2019YFA0405500, National Natural Science Foundation of China 12173034, 12322304, 12222301 and 12173007, and Yunnan University grant No.~C619300A034. We acknowledge the science research grants from the China Manned Space Project with NO.\,CMS-CSST-2021-A09, CMS-CSST-2021-A08 and CMS-CSST-2021-B03. 

This work is based on data products from observations made with ESO Telescopes at the La Silla Paranal Observatory under programme ID 177.D-3023, as part of the VST Photometric H$\alpha$ Survey of the Southern Galactic Plane and Bulge (VPHAS+, www.vphas.eu).

This work has made use of data from the European Space Agency (ESA) mission {\it Gaia} (\url{https://www.cosmos.esa.int/gaia}), processed by the {\it Gaia} Data Processing and Analysis Consortium (DPAC, \url{https://www.cosmos.esa.int/web/gaia/dpac/consortium}). Funding for the DPAC has been provided by national institutions, in particular the institutions participating in the {\it Gaia} Multilateral Agreement.

\section*{Data availability}

The data underlying this article is available in the manuscript.

\bibliographystyle{mnras}
\bibliography{vphasu} 








\bsp	
\label{lastpage}
\end{document}